\documentclass[11pt]{article}
\pdfoutput=1
\usepackage{jheppub}
\usepackage{amssymb,amsfonts}
\usepackage{mathrsfs}
\let\savenumberline\numberline
\def\numberline#1{\savenumberline{#1.}}
\makeatletter
\renewcommand{\@seccntformat}[1]{\csname the#1\endcsname.\,\,}
\makeatother

\newcommand{\CF}{{\cal F}}

\newcommand{\CL}{{\cal L}}

\newcommand{\CN}{{\cal N}}

\newcommand{\CZ}{{\cal Z}}

\newcommand{\SC}{{\mathscr C}}

\newcommand{\p}{\partial}
\renewcommand{\tilde}[1]{\widetilde{#1}}
\renewcommand{\hat}[1]{\widehat{#1}}
\newcommand{\be}{\begin{equation}}
\newcommand{\ee}{\end{equation}}
\newcommand{\bea}{\begin{eqnarray}}
\newcommand{\eea}{\end{eqnarray}}

\newcommand{\ie}{\textit{i.e.}}
\newcommand{\eg}{\textit{e.g.}}

\newcommand{\im}{\text{Im}}

\newcommand{\Tr}{\textrm{Tr}}
\newcommand{\sigp}{{\Sigma^+}}
\newcommand{\sigm}{{\Sigma^-}}
\newcommand{\sigw}{{\Sigma^\wedge}}

\newcommand{\rmm}{\textrm{M}}
\newcommand{\T}{\textrm{T}}
\makeatletter
\def\@fpheader{\relax}
\makeatother
\usepackage{graphicx}
\usepackage{latexsym}

\title{\ \vspace{1.5in} \\ \hbox{Large-N Expansion and String Theory Out of Equilibrium}}
\author{Petr Ho\v{r}ava and Christopher J. Mogni}
\affiliation{Berkeley Center for Theoretical Physics and Department of Physics\\
University of California, Berkeley, CA, 94720-7300, USA\medskip\\
Theoretical Physics Group, Lawrence Berkeley National Laboratory\\
Berkeley, CA 94720-8162, USA}
\abstract{We analyze the large-$N$ expansion of general non-equilibrium systems with fluctuating matrix degrees of freedom and $SU(N)$ symmetry, using the Schwinger-Keldysh formalism and its closed real-time contour with a forward and backward component.  In equilibrium, the large-$N$ expansion of such systems leads to a sum over topologies of two-dimensional surfaces of increasing topological complexity, predicting the possibility of a dual description in terms of string theory.  We extend this argument away from equilibrium, and study the universal features of the topological expansion in the dual string theory.  We conclude that in non-equilibrium string perturbation theory, the sum over worldsheet topologies is further refined:  Each worldsheet surface $\Sigma$ undergoes a triple decomposition into the part $\sigp$ corresponding to the forward branch of the time contour, the part $\sigm$ on the backward branch, and the part $\sigw$ that corresponds to the instant in the far future where the two branches of the time contour meet.  The sum over topologies becomes a sum over the triple decompositions.  We generalize our findings to the Kadanoff-Baym time contour relevant for systems at finite temperature, and to the case of closed and open, oriented or unoriented strings.  Our results are universal, and follow solely from the features of the large-$N$ expansion without any assumptions about the worldsheet dynamics.}
\begin{document}
\maketitle
\section{Introduction}

Our universe is not in equilibrium.%
\footnote{See, \eg , \cite{glass}.}
The framework of string theory has successfully provided a consistent theoretical picture for describing various aspects of its dynamics, capable of accommodating both the quantum mechanical nature of its constituents and the evolving geometry of its large-scale structure.  Yet, somewhat paradoxically, the machinery of string theory as understood today does not appear to be particularly well-suited for describing systems out of equilibrium, such as early-universe cosmology.

Concepts originating from string theory have been very influential in a remarkable number of areas of physics (and even mathematics).  This interdisciplinary influence of string theory includes particle phenomenology, with brane-world scenarios, large extra dimensions, the Randall-Sundrum scenario enriching the scene beyond the Standard Model; AdS/CMT and holographic methods for describing strongly-correlated condensed matter systems  \cite{zaanen,hartnoll}; the extension of K-theory from a method for classifying D-branes in string theory to classifying stable Fermi surfaces \cite{kth} and phases of topological insulators; and the impact of string theory on inflationary cosmology \cite{baumann} and in quantum gravity, notably leading to the statistical explanation of the Bekenstein-Hawking entropy of various supersymmetric black holes.    

In most of these applications, string theory is excellent at describing equilibrium systems, ideally with as many supersymmetries as possible.  However, this effectiveness seems to be lost for systems or states away from equilibrium.  One naturally wonders why:  Is this a fundamental limitation of string theory?  Or is it a historical accident, with the proper formulation of string theory away from equilibrium yet to be discovered?  Indeed, a glance at the history of string theory reveals a strong bias towards equilibrium states.  Since its inception in the 1960's and certainly for much of its early development \cite{birth}, string theory has been deeply rooted in the ideology of the S-matrix, which depends strongly on the axiom of a static, stable, eternal vacuum.

Can we uncouple string theory from this assumption of the eternal stable vacuum?  
While many partial results for string-theory states away from equilibrium have been accumulated -- notably, in areas ranging from tachyon condensation to non-equilibrium AdS/CFT dynamics --  progress has been rather slow and spotty.  It is natural to hope that even in its natural area of quantum gravity, string theory should be able to do better with non-equilibrium systems, to have a more systematic impact of string theory on concepts in early-universe cosmology, or to give new insights into dynamical evaporating black holes.  

It may not be immediately clear how to wean critical string theory from its dependence on the S-matrix and equilibrium, or how to formulate non-equilibrium string theory from first principles.  However, we do know how to take a general quantum many-body system or quantum field theory out of equilibrium:  The basic rules of quantum mechanics can certainly accommodate non-equilibrium states, leading to the formulation known as the Schwinger-Keldysh formalism.  In fact, in recent years the methods of the Schwinger-Keldysh formalism have found their way into string theory, primarily in the context of AdS/CFT \cite{sonh,kostas1,kostas2,haehl1,haehl2,jandb}.  However, these approaches are mostly based on the spacetime field theory description, with very little understanding so far of the worldsheet dynamics.  

Here we will follow a different strategy:  We use the methods of the large-$N$ expansion and its connection to string theory, and extend them to non-equilibrium systems where we can directly apply the Schwinger-Keldysh formalism.  In that way, we begin to learn something about the universal rules of non-equilibrium string perturbation theory.  Our goal is two-fold:  To stimulate string-theory research in directions away from equilibrium, and to encourage further study of possible dual descriptions of non-equilibrium systems across diverse areas of physics in terms of string theory.  

This paper is organized as follows.  In the remainder of this introductory Section~1, we briefly review two important topics: The interpretation of the large-$N$ expansion in a quantum theory of matrices in terms of string theory, and the non-equilibrium formalism for quantum systems known as the Schwinger-Keldysh formalism.  The reviewed material is well-known to experts in the corresponding fields, but since we wish to make this paper accessible to a broad audience from a wide range of fields -- from string theory to non-equilibrium mesoscopic physics to early-universe cosmology -- we include this material to make our paper relatively self-contained, and to set a uniform stage for the later sections.  In Section~2, we connect the two topics reviewed in Section~1, and analyze how the large-$N$ expansion of the non-equilibrium Schwinger-Keldysh formalism leads to a refined expansion in terms of string theory topologies.  It is the hallmark of the Schwinger-Keldysh formalism that the system is followed forward and then backward in time, and we analyze how string perturbation theory out of equilibrium reflects this doubling phenomenon.  We concentrate on aspects which are universal, and follow solely from the structure of the large-$N$ expansion; we make no assumptions about worldsheet dynamics.   We perform our analysis for the case of matrices with $SU(N)$ symmetry, which corresponds to the case of closed oriented strings. We develop the universal structure of non-equilibrium string perturbation theory in terms of a refined sum over worldsheet topologies.

Sections~3 and 4 are then devoted to several generalizations of our main results from Section~2.  In Section~3, we consider an important special case, particularly useful for studies of equilibrium systems at finite temperature $T$.  Here the relevant time contour -- often referred to as the Kadanoff-Baym contour -- contains not only the forward and backward evolution segments in real time familiar from the Schwinger-Keldysh contour, but also a ``Matsubara segment''along the imaginary time direction by the amount $\beta=1/T$.  This approach naturally contains both the real-time and imaginary-time approaches to systems at nonzero $T$.  We analyze how the large-$N$ theory on the Kadanoff-Baym contour leads to a further refinement of the expected universal features of string perturbation theory.  In Section~4, we briefly outline the generalizations of our main results from Section~2 to the case of matrices with $SO(N)$ or $Sp(N)$ symmetries, which lead to closed unoriented strings; and the addition of vector-like degrees of freedom, in the fundamental representation of the appropriate symmetry group, which leads to open strings and the presence of worldsheet boundaries.  We conclude in Section~5.

A brief summary of our results (without proofs) appears in the short companion paper \cite{ssk}, which also contains additional results on the topological expansion in the Keldysh-rotated version of the Schwinger-Keldysh formalism.

\subsection{Strings from the large-\textit{N} expansion}
\label{ssnstr}

The genus expansion into worldsheets of increasing topological complexity, weighted by the powers of the string coupling $g_s$, is a universal hallmark of string theory in its perturbative regime.  It is remarkable that the same topological expansion is obtained, quite universally, in the large $N$ limit of theories with degrees of freedom described by matrices of rank $N$, with $1/N$ playing the role of the string coupling constant $g_s$. The large-$N$ expansion has turned into an efficient strategy for reorganizing theories that would otherwise be difficult to understand perturbatively.  In the context of high-energy physics, the use of this strategy to illuminate QCD dynamics goes back to 1974 and G.~'t~Hooft \cite{th,th2,thspell}.  Quite universally, the large-$N$ expansion predicts the existence of a dual description of the same system in terms of string theory.  This association with the large-$N$ description of generic systems of fluctuating matrix degrees of freedom is one of the most compelling arguments for the importance of string theory.  For readable reviews of the elements of the large-$N$ approach, see \cite{colemann} (reprinted in \cite{coleman}), or the more recent \cite{juantasi}.

We begin with a system of fluctuating degrees of freedom, described by $M$ which happens to be an $N\times N$ matrix, which we take to be Hermitian and traceless, so that it carries the adjoint representation of our symmetry group $SU(N)$.  This matrix may depend on spacetime coordinates, and its dynamics may be relativistic or not; the details are immaterial, and we suppress them in what follows.  We will study the system in the perturbative expansion in the powers of $1/N$.  The limit of large $N$ will correspond to a new classical limit \cite{yaffe}, in a dual theory described by strings.  For simplicity, we assume that the system is defined by a path integral, with a classical action $S(M)$.  $M$ can be relativistic Yang-Mills gauge fields%
\footnote{If $M$ are Yang-Mills gauge fields or if there is any other gauge symmetry, we assume that the gauge symmetry is handled in the BRST formalism, extending the matrix degrees of freedom to include ghosts and antighosts such that each matrix field has a non-degenerate kinetic term and a well-defined propagator, so that the ribbon-diagram expansion discussed below makes sense.}%
, or they can be nonrelativistic matrix fields in some number of spatial dimensions.  They can also just be $N\times N$ matrices in quantum mechanics, dependent only on time.  The beauty of the large-$N$ expansion argument that we are about to review is in its universality.  

In order to set the stage for our arguments, we must choose an action for $M$.  We will mimic the case of Yang-Mills gauge theory, and will take the action to be
\be
\label{eeaction}
S(M)=\frac{1}{2g^2}\int dt\,\Tr\left(\dot M^2+M^3+M^4+\ldots\right).
\ee
In the quadratic term, we indicated explicitly only the piece with time derivatives, but generally there will also be terms involving spatial derivatives, as well as mass/chemical potential terms; we keep those implicit, focusing on the universal features only.  The propagator is determined by the full quadratic part in $M$.

A simple field redefinition to $m=M/g$ would take this action to another, perhaps more familiar form, traditionally used for perturbation theory in $g$:
\be
S(m)=\int dt\,\Tr\left(\frac{1}{2}\dot m^2+ \frac{g}{2} m^3+\frac{g^2}{2}m^4+\ldots\right).
\ee
Here the quadratic term is normalized to $1/2$, and each interaction term is controlled by the appropriate power of $g$.  It is also not difficult to generalize this and make the $M^4$ coupling constant independent of the coupling that controls the $M^3$ term.  Such cosmetic modifications will not change the our line of reasoning.  Importantly, the change from $M$ to $m$ is just a simple change of coordinates, which will not influence the underlying physics.  We feel that our arguments will be simplest in the original notation using $M$, and will use that parametrization in the rest of our analysis.

The propagator defined by the full quadratic part of the action in (\ref{eeaction}) is depicted by a ribbon, with each of the two indices associated with one edge of the ribbon,
\be
\vcenter{\hbox{\includegraphics[width=.9in]{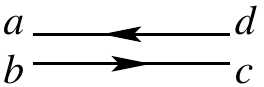}}}\ =\left\langle M^a_{\ b}M^c_{\ d}\right\rangle=g^2\,G{}^{ac}_{\ bd}=g^2\,G\,\delta^a_d\delta^c_b.
\label{eenprop}
\ee
The arrows at the edges distinguish the upper and lower indices.%
\footnote{\label{ftnt}A standard word of explanation and caution about the distinction between $U(N)$ and $SU(N)$:  By our assumptions, the $M$ degrees of freedom are traceless, and the symmetry is $SU(N)$.  The correct propagator would then contain also an additive term $-(1/N)\delta^a{}_b\delta^c{}_d$ on the right-hand side of (\ref{eenprop}), in order to maintain the tracelessness of $M$.  We drop this terms systematically in the large-$N$ expansion.  Thus, we aproximate $SU(N)$ by $U(N)$, which is permissible as long as the $U(1)$ factor is free and decouples (which we assume throughout this paper).  For further discussion of this standard approach, see \cite{coleman}.}
The bare propagator $G$ can be a function of various suppressed arguments of $M$, but is independent of $g$.  The vertices are
\bea
\vcenter{\hbox{\includegraphics[width=.75in]{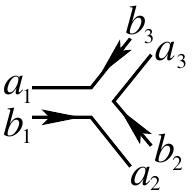}}}\ \ \ &=&\ \ -\frac{i}{g^2}\,\delta^{a_2}_{b_1}\delta^{a_3}_{b_2}\delta^{a_1}_{b_3},\\
\vcenter{\hbox{\includegraphics[width=.8in]{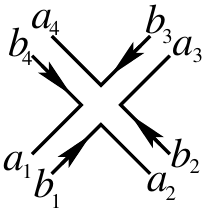}}}\ &=&\ \ -\frac{i}{g^2}\,\delta^{a_2}_{b_1}\delta^{a_3}_{b_2}\delta^{a_4}_{b_3}\delta^{a_1}_{b_4},\\
\vdots&&
\eea
Feynman diagrams built from these propagators and vertices are often called ``ribbon diagrams'', and this is the terminology we will use in this paper.%
\footnote{Historically, ribbon diagrams appeared independently in the mathematical literature, where they are often referred to as ``fatgraphs'' (see, \eg, \cite{penner} and references therein).}
Let us focus for simplicity on vacuum ribbon diagrams. For a generic ribbon diagram, we will denote by $P$ its number of propagators, by $V$ the number of vertices and by $L$ the number of closed loops.  We will also denote the ribbon diagram itself by $\Delta$.

Each ribbon diagram $\Delta$ can be uniquely associated with a compact surface $\Sigma$.   Loosely speaking, $\Sigma$ is the lowest-genus surface on which the ribbon diagram can be drawn.  More precisely, the constructive prescription for obtaining this $\Sigma$ for a given ribbon diagram is very simple: Start with the ribbon diagram (as a topological 2-manifold, with boundaries consisting of the edges of the ribbons), and for each closed loop (\ie, a boundary component which is topologically an $S^1$) glue in a two-dimensional disk $D_2$, thus closing all holes in the ribbon diagram and producing a compact surface $\Sigma$ with $\p\Sigma=\emptyset$.  In turn, the ribbon diagram gives a cellular decomposition of $\Sigma$, with the vertices and propagators of the diagram serving as the 0-dimensional and 1-dimensional cells, while the glued-in disks -- which we will refer to as ``plaquettes'' -- play the role of the 2-dimensional cells in this cellular decomposition of $\Sigma$.  When we wish to indicate explicitly which ribbon diagram $\Delta$ gave rise to a given surface, we will denote that surface by $\Sigma(\Delta)$.  

\begin{figure}[t!]
  \centering
    \includegraphics[width=0.24\textwidth]{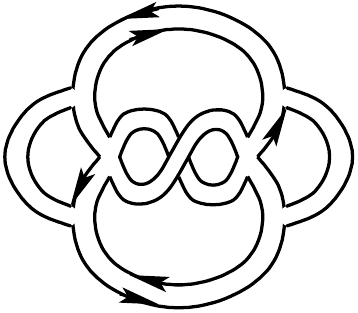}
    \caption{A typical ribbon diagram, with 6 vertices, 10 propagators, and 4 closed loops.  The Riemann surface associated with this diagram has Euler number $\chi(\Sigma)=V-P+L=0$, \ie, it is the surface of genus one, $\Sigma=T^2$.}
    \label{ffribbon}
\end{figure}

By Feynman rules, the contribution of a given ribbon diagram to the vacuum amplitude depends on $g$ and $N$ as
\be
g^{2P-2V}N^L.
\ee
Importantly, the factor of $N^L$ appears because each plaquette corresponds to a closed loop, and therefore includes the summation over the $N$ values of the index $a$ running around the loop.

We are primarily interested in a meaningful $1/N$ expansion, and therefore have to determine which combination of $g$ and $N$ to hold fixed as $N\to\infty$ in order to achieve this.  Defining the 't~Hooft coupling $\lambda$
\be
\lambda\equiv g^2\, N
\ee
turns this scaling to
\be
\lambda^{2P-2V}N^{V-P+L}.
\ee
We recognize the power of $N$ in this expression as
\be
\label{eeeuler}
\chi(\Sigma)\equiv V-P+L,
\ee
the Euler number $\chi(\Sigma)$ of the surface $\Sigma$ associated to the ribbon diagram by the construction summarized above.  In (\ref{eeeuler}), $\chi(\Sigma)$ is expressed in terms of the combinatorial data about $\Sigma$.  It is crucial, however, that $\chi(\Sigma)$ is a topological invariant of $\Sigma$, independent of the specific cellular decomposition of $\Sigma$ into a collection of vertices, lines and plaquettes.

Famously, topologically inequivalent compact oriented Riemann surfaces are fully classified by specifying just one non-negative integer, the genus $h$ of the surface, and we have $\chi(\Sigma)=2-2h$.  Hence, our $1/N$ expansion is naturally interpreted as organized according to the increasing complexity of the topology of $\Sigma$.  All diagrams can now be resummed into a perturbative expansion in the powers of $1/N$, and the partition function can be written as
\be
\CZ=\sum_{h=0}^\infty \left(\frac{1}{N}\right)^{2h-2}\CF_h(\lambda,\ldots).
\ee
We define the large-$N$ limit by holding the 't~Hooft coupling fixed, and identify $1/N$ as the string coupling constant,
\be
g_s=\frac{1}{N}.
\ee
We showed the analysis for simplicity for the partition function, but the same conclusion extends to the correlation functions of physical observables in the underlying theory of the matrix degrees of freedom:  There is a dual interpretation of this theory as a string theory.

This argument is very convincing in its generality and universality.  The catch with this simple universal argument is that it does not give us \textit{a priori} clues as to which string theory is dual to our system.  The worldsheet dynamics of the string needs to be found by other independent means, which are available only in a few rare cases (such as maximally supersymmetric Yang-Mills theories whose additional features allow the dual string theory to be uniquely determined, leading to the celebrated AdS/CFT correspondence \cite{juan}).  

\subsection{Quantum theory in real time: Schwinger-Keldysh formalism}

The relationship between the large-$N$ expansion and a perturbative string-theory expansion as reviwed in Section~\ref{ssnstr} is derived under a very important implicit assumption, with historical roots in particle physics: The assumption that the system is in a stable, eternal, static vacuum, or in a state not too far from it.  Our main goal in this paper is to relax this assumption, and study the large-$N$ expansion away from equilibrium.  Such systems are naturally described by a natural generalization of standard quantum field theory, known as the Schwinger-Keldysh formalism.

Here we give a lightning review of Schwinger-Keldysh formalism, which describes quantum theory for general states, in or out of equilibrium \cite{schwinger,keldysh}.  There are many useful reviews of this formalism, scattered across various fields of physics; see, \eg , \cite{ll,rammers,rammer,mahan,chou,lebellac,das,dast,kamenev,svl,gelis,vilkovisky,spicka1,spicka2,spicka3,spicka}.  Schwinger-Keldysh formalism is also sometimes referred to as the ``in-in'' formalism \cite{vilkovisky}, especially in cosmology \cite{weinberg1,weinberg2,weinberg3}.%
\footnote{Although the in-in formalism is the consequence of the same quantum mechanics as the Schwinger-Keldysh formalism, it might be appropriate to point out that the physical focus is a bit different:  In the cosmological in-in formalism, one concentrates on the correlation functions of observables located at $t_\wedge$, which is interpreted as ``the present''.  In the Schwinger-Keldysh formalism, $t_\wedge$ represents ``the end of time'' in the future, and the correlators are typically evaluated for observables on the forward time contour before $t_\wedge$ is reached.} 
All these labels for this formalism are largely historical; it would be sensible to think of this formalism simply as ``quantum mechanics without simplifying assumptions about the vacuum''.

\begin{figure}[t!]
  \centering
    \includegraphics[width=0.12\textwidth]{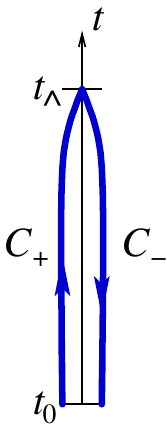}
    \caption{Schwinger-Keldysh closed time contour $\SC=C_+\cup C_-$.  The remote past $t_0$ and the far future $t_\wedge$ are usually taken to be $-\infty$ and $+\infty$.} 
    \label{ffctc}
\end{figure}

The main highlight of the Schwinger-Keldysh formalism is that it describes the system as evolving on a doubled closed-time contour $\SC$ (known as the Schwinger-Keldysh time contour, see Fig.~\ref{ffctc}), starting in the remote past, evolving to the far future $t_\wedge$ along the forward part $C_+$ of the time contour, and then returning along the backward part $C_-$ of the contour back to the remote past.  Often the turn-around point is taken $t_\wedge\to\infty$.  

Why such a closed time contour?  In fact, this contour is encoded automatically in the rules of quantum mechanics, if one does not make the simplifying assumption of the static vacuum.  To see this, let us focus on simple observables:  Time-ordered correlation functions of operators in the Heisenberg picture, 
\be
\left\langle\psi_{\textrm{in}}|\T(\phi_{\textrm{H}}(t_n)\ldots\phi_{\textrm{H}}(t_1))|\psi_{\textrm{in}}\right\rangle,
\label{eecorr}
\ee
in some general initially prepared state $\left|\psi{_\textrm{in}}\right\rangle$.  If this state is the static, stable vacuum, the standard LSZ procedure extracts from these correlators the physically observable S-matrix elements.  Those are also the natural observables in string theory.  

If $|\psi{_\textrm{in}}\rangle$ is not the static, stable vacuum, we can still apply standard rules of quantum mechanics and develop a perturbative expansion for (\ref{eecorr}).  Assume that the Hamiltonian can be written as
\be
H=H_0+V(t),
\ee
where $H_0$ describes a simple system, and define the interaction picture using this split.  The interaction-picture operators $\phi(t)$ are related to the Heisenberg-picture operators $\phi_{\textrm{H}}$ by 
\be
\phi_{\textrm{H}}(t)=\textrm{S}(t_0,t)\,\phi(t)\,\textrm{S}(t,t_0).
\label{eepsih}
\ee
Here $\textrm{S}(t',t)$ is the evolution operator 
\be
\textrm{S}(t',t)=\T\exp\left(-i\int_t^{t'} V_0(t'')\,dt''\right),
\label{ees}
\ee
and $V_0(t)$ is the interaction part $V(t)$ of the Hamiltonian in the interaction picture.
We will denote $\textrm{S}(+\infty,-\infty)$ simply by $\textrm{S}$.

The fixed reference time $t_0$ in (\ref{eepsih}) can be taken to be in the remote past.  
We may also assume, for illustration, that $|\psi_{\textrm{in}}\rangle$ was prepared from the vacuum $|0_{\textrm{in}}\rangle$ of $H_0$ in the remote past, by adiabatic turning-on of the iteractions.  The correlators are then
\be
\left\langle 0_{\textrm{in}}|\textrm{S}^{-1}\,\T(\textrm{S}\,\phi(t_s)\ldots \phi(t_1)|0_{\textrm{in}}\right\rangle.
\label{eekscorr}
\ee
Note that the factor of $\textrm{S}^{-1}$ is automatically present, and it serves to evolve the system back from the infinite future to the remote past where $\langle 0_{\textrm{in}}|$ was prepared:
\be
\textrm{S}^{-1}=\left[\textrm{S}(+\infty,-\infty)\right]^{-1}=\textrm{S}(-\infty,+\infty).
\ee
Clearly, the perturbative expansion of (\ref{eekscorr}) will involve not just time-ordered two-point functions of $\phi$, but also anti-chronologically ordered ones, and unordered ones as well.  This proliferation of propagators is best encoded by defining the closed time contour $\SC$, with the factor of $\textrm{S}$ under the time ordering symbol $\T$ in (\ref{eekscorr}) evolving the system forward in time along $C_+$, and the factor of $\textrm{S}^{-1}$ outside of $\T$ evolving back along $C_-$.  We introduce the time-ordering symbol $\T_\SC$ to denote chronological ordering along the entire contour, allowing (\ref{eekscorr}) to be succinctly written as
\be
\left\langle 0_{\textrm{in}}|\T_\SC(\textrm{S}_\SC\,\phi(t_s)\ldots \phi(t_1))|0_{\textrm{in}}\right\rangle,
\ee
with $\textrm{S}_\SC$ the evolution operator (\ref{ees}) along the entire contour $\SC$.  Only when the final vacuum $|0_{\textrm{fin}}\rangle$ is given by the initial vacuum up to a possible phase,
\be
|0_{\textrm{fin}}\rangle=e^{i\theta}|0_{\textrm{in}}\rangle,
\ee
can we replace $\langle 0_{\textrm{in}}|\textrm{S}^{-1}$ by $e^{i\theta}\langle 0_{\textrm{fin}}|$ and obtain the standard perturbation theory involving only the Feynman propagators of $\phi$.  In more general circumstances, however, we cannot replace the initial state $\langle 0_{\textrm{in}}|$ with a suitable out-state, simply because the final state is not known.  We must then follow the general formula (\ref{eekscorr}) and evolve the system back using $\textrm{S}^{-1}$, before closing the correlator on the known initial state.

It is often impractical to work directly with the doubled time contour $\SC$.  Instead, one can keep the single-valued time $t$, and double the number of fields, with $\phi_+(t)$ and $\phi_-(t)$ denoting $\phi(t)$ on the $C_+$ and $C_-$ branch of the Schwinger-Keldysh contour $\SC$ at the same value of $t$.  These doubled fields can be used in the path integral representation of the theory.  The action that appears in the path integral of the non-equilibrium system is then formally given by
\be
S_{\textrm{SK}}=\int_{-\infty}^{+\infty}dt\, \left\{\CL(\phi_+)-\CL(\phi_-)\right\},
\label{ssactsk}
\ee
with $S=\int\CL(\phi)$ the original action of the equilibrium system.  Note, however, that the compact form (\ref{ssactsk}) is somewhat deceiving, and careful arguments involving regulators may be needed to provide the correct treatment of the non-equilibrium path integral (see \cite{kamenev} for details).

\section{Large-\textit{N} expansion in quantum systems out of equilibrium}
\label{ssnneq}

In this section, we put the Schwinger-Keldysh formulation of non-equilibrium systems together with the large-$N$ expansion, and analyze the consequences of the Schwinger-Keldysh formalism for string perturbation theory.  In particular, we wish to understand how the Schwinger-Keldysh time contour is perceived by the string worldsheet topologies.

\subsection{Ribbon diagrams on the Schwinger-Keldysh time contour}

First, we formulate Feynman rules out of equilibrium, for our theory of Hermitian traceless matrices $M^a{}_b$, in the adjoint of $SU(N)$.  The elements of Feynman graphs again lead to ribbon diagrams, but now with all vertices and all ends of propagators labeled with $+$ or $-$.  The propagators are:
\bea
\vcenter{\hbox{\includegraphics[width=1in]{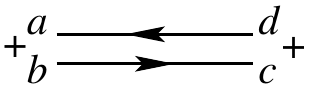}}}\ &=&\left\langle\T_\SC\left(M^a_{+b}M^c_{+d}\right)\right\rangle=g^2\,G_{++}{}^{ac}_{\ bd},
\label{eeproppp}
\\
\vcenter{\hbox{\includegraphics[width=1in]{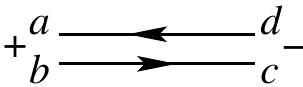}}}\ &=&\left\langle\T_\SC\left( M^a_{+b}M^c_{-d}\right)\right\rangle=g^2\,G_{+-}{}^{ac}_{\ bd},
\label{eeproppm}
\\
\vcenter{\hbox{\includegraphics[width=1in]{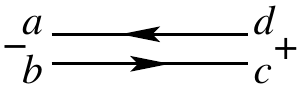}}}\ &=&\left\langle\T_\SC\left( M^a_{-b}M^c_{+d}\right)\right\rangle=g^2\,G_{-+}{}^{ac}_{\ bd},
\label{eepropmp}
\\
\vcenter{\hbox{\includegraphics[width=1in]{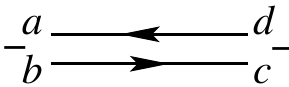}}}\ &=&\left\langle\T_\SC\left( M^{a}_{-b}M^c_{-d}\right)\right\rangle=g^2\,G_{--}{}^{ac}_{\ bd}.
\label{eepropmm}
\eea
The operation $\T_\SC$ of time-ordering along the contour $\SC$ again acts on its arguments by reordering them from the right to the left in the order of increasing contour time, with the flow of time following the direction of the arrows on the contour in Fig.~\ref{ffctc}.

More explicitly, the $\T_\SC$ time ordering can be understood in terms of the more elementary orderings on the standard time axis parametrized by coordinate time $t$:  The chronological time ordering $\T$ along $t$, and the anti-chronological ordering $\overline\T$, in the reverse direction of $t$.  Here we suppress the ${}^{ab}{}_{cd}$ indices for simplicity, but restore the time dependence, while still suppressing the spatial dependence and all other possible indices and quantum numbers of $M_\pm$:
\bea
\left\langle\T_\SC\left(M_+(t)M_+(t')\right)\right\rangle
&=&\left\langle\T\left(M(t)M(t')\right)\right\rangle=g^2\,G_F(t,t'),\label{eetimepp}\\
\left\langle\T_\SC\left( M_+(t)M_-(t')\right)\right\rangle
&=&\left\langle M(t')M(t)\right\rangle=g^2\,G^<(t,t'),\\
\left\langle\T_\SC\left( M_-(t)M_+(t')\right)\right\rangle
&=&\left\langle M(t)M(t')\right\rangle=g^2\,G^>(t,t'),\\
\left\langle\T_\SC\left( M_-(t)M_-(t')\right)\right\rangle
&=&\left\langle\overline{\T}\left( M(t)M(t')\right)\right\rangle=g^2\,G_{\overline{F}}(t,t').
\label{eetimemm}
\eea
We thus recognize all four types of propagators in (\ref{eeproppp}-\ref{eepropmm}) in more elementary terms, as representing the Feynman $i\varepsilon$ propagator $G_F$, the ``anti-Feynman'' propagator $G_{\overline{F}}$ (sometimes called the Dyson propagator), and the $G$-lesser and $G$-greater propagators $G^<$, $G^>$.  For clarity and simplicity, we will keep our $G_{\pm\pm}$ notation of (\ref{eeproppp}-\ref{eepropmm}) throughout the paper.

The vertices look the same as in the equilibrium case, except that each vertex is assigned a $\pm$ sign:
\bea
\vcenter{\hbox{\includegraphics[width=.75in]{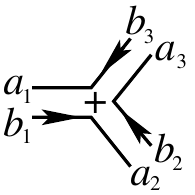}}}\ \ \ &=&\ \ -\frac{i}{g^2}\,\delta^{a_2}_{b_1}\delta^{a_3}_{b_2}\delta^{a_1}_{b_3},
\label{eevert+}
\\
\vcenter{\hbox{\includegraphics[width=.75in]{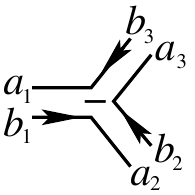}}}\ \ \ &=&\ \ \frac{i}{g^2}\,\delta^{a_2}_{b_1}\delta^{a_3}_{b_2}\delta^{a_1}_{b_3},\\
\vcenter{\hbox{\includegraphics[width=.8in]{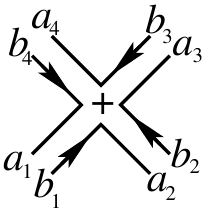}}}\ &=&\ \ -\frac{i}{g^2}\,\delta^{a_2}_{b_1}\delta^{a_3}_{b_2}\delta^{a_4}_{b_3}\delta^{a_1}_{b_4},\\
\vcenter{\hbox{\includegraphics[width=.8in]{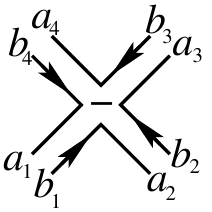}}}\ &=&\ \ \frac{i}{g^2}\,\delta^{a_2}_{b_1}\delta^{a_3}_{b_2}\delta^{a_4}_{b_3}\delta^{a_1}_{b_4},
\label{eevert--}\\
\vdots &&\nonumber
\eea
The vertical dots at the bottom of this list of vertices represent the possibility of having higher, $n$-point vertices beyond $n=4$.  These higher-point vertices can be controlled by $1/g^2$, or they can have their own independent couplings.  These additional choices do not change the universal results of our analysis, and we will often freely assume below, simply for convenience, that such higher-$n$ vertices do exist.  Similarly, we assume for simplicity that all vertices have at least three ends; the arguments could be easily extended if one added ``2-vertices'' and ``1-vertices'' as well, without altering our conclusions.

Note that because of the Hermiticity of $M$ and the nature of the time ordering along $\SC$, we have
\be
\label{eesame}
G_{+-}{}^{ac}_{\ bd}=G_{-+}{}^{ca}_{\ db},
\ee
and there is therefore only one independent propagator that can connect a $+$ vertex to a $-$ vertex.  This is reflected in our graphical notation:  
The propagator on the left side of (\ref{eeproppm}), after the ends of the ribbon are exchanged and the indiced swapped, looks identical to the propagator on the left side of (\ref{eepropmp}).  This means that we will not have to distinguish between $G_{+-}$ and $G_{-+}$ propagators, as long as they are attached to the apprpriate $+$ and $-$ vertices.  As a result, each ribbon diagram in non-equilibrium perturbation theory will look like a ribbon diagram of the type we encountered in Section~\ref{ssnstr} at equilibrium, but now with each vertex labeled by a $\pm$ sign.%
\footnote{In what follows, when we draw ribbon diagrams we will often put the $\pm$ sign next to the vertex rather than inside the ribbon; this will make some of our diagrams easier to read.}

Now we proceed to the analysis of generic ribbon diagrams, and we study how they lead to an expansion of the partition function and correlation functions in terms of the topology of surfaces, generalizing the well-known string perturbation theory away from equilibrium.  We will refer to the surfaces representing string worldsheets as ``Riemann surfaces'' for short, without implying that any geometric structure on them is \textit{a priori} assumed, besides their smooth manifold structure.

\subsection{First look at string perturbation theory out of equilibrium}

In the special case of equilibrium and zero temperature, the Schwinger-Keldysh formalism should correctly reproduce the standard formulation of equilibrium quantum field theory in real time $t$.  This limit is usually taken such that as we sent $t_\wedge\to\infty$, the return part of the Schwinger-Keldysh contour decouples from the calculations of the correlation functions of operators located on the forward branch, and therefore it can be ignored, reproducing standard textbook rules of quantum field theory with the static eternal vacuum.  However, remnants of the Schwinger-Keldysh formalism do appear even in this textbook example of vacuum correlation functions in equilibrium at zero temperature, in an almost clandestine way, under a very different name:  it reduces to the Cutkosky rules, which are crucial for analyzing unitarity properties of physical amplitudes \cite{diagrr,veltmanun,veltman}.%
\footnote{This remarkable connection between the Schwinger-Keldysh formalism and the Cutkosky rules seems absent in most textbooks on relativistic quantum field theory.  One notable exception, where this relationship is explained in a lucid way, is the recent textbook by Gelis \cite{gelis}.}
Indeed, we can take the $t_\wedge\to\infty$ limit for the vacuum correlators in equilibrium at zero temperature, but still allow insertions of observables along both branches of $\SC$.  The $-$ vertices and operator insertions located on the backward branch $C_-$ behave exactly like those on the ``shaded side'' from the unitarity cuts.  Similarly, the propagators on the ``unshaded'' or ``shaded'' sides are simply the equilibrium limits of $G_{++}$ and $G_{--}$, and the ``cut propagators'' of the Cutkosky formalism correspond to the equilibrium limit of $G_{+-}$, where this propagator reduces to the on-shell delta function.  Thus, we reproduce the standard Cutkosky rules from the Schwinger-Keldysh formalism: The ``shaded'' and ``unshaded'' portions in the Cutkosky rules for Feynman diagrams correspond to the forward and backward branch of the Schwinger-Keldysh time contour, and the cut between the shaded and unshaded region is simply the location of the crossing from the forward branch $C_+$ to the backward branch $C_-$, at $t_\wedge\to\infty$.  

We now wish to extend the story of the large-$N$ expansion and string theory away from equilibrium.  The first guess might be that propagating strings will also exhibit cuts, and that each string worldsheet $\Sigma$ will consequently be split into two parts -- its forward and backward portions $\sigp$ and $\sigm$, joined along a shared one-dimensional boundary $\p\sigp=\p\sigm$.  This common boundary between $\sigp$ and $\sigm$ would then represent the cuts in the worldsheet language.  It is one of the central points of this paper to show that such an expectation is not quite correct.  Instead, we will find that the portion of the worldsheet connecting $\sigp$ and $\sigm$ is topologically two-dimensional.

\subsection{Extending the cuts}

Intuitively, the propagators that connect a vertex on the forward portion $C_+$ of the time contour with a vertex on the $C_-$ portion of the contour represent worldlines of particles that have to cross from $C_+$ to $C_-$ and therefore pass through the time instant $t_\wedge$ where the two branches meet.  This crossing can be usefully denoted in Feynman diagrams by placing cuts across such propagators, indicating the passage through $t_\wedge$.   This suggests that in the string picture, such cuts should be perhaps extended from cuts of ribbon diagrams to worldsheet cuts.

Let us first  test this guess by considering some simple examples of ribbon diagrams.  We begin by placing a cut line across all $G_{+-}$ and $G_{-+}$ propagators,
$$
\vcenter{\hbox{\includegraphics[width=1in]{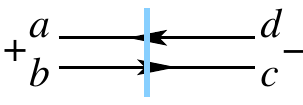}}}\ \ \ .
$$
Intuitively, one can think of the cut as indicating where the worldline of the virtual particle, represented by the propagator, crosses over from the forward branch $C_+$ to the backward branch $C_-$ of the Schwinger-Keldysh time contour $\SC$, where the two ends of the propagator are located.  If our expectation about cuts of surfaces were correct, such cuts on ribbon propagators should induce uniquely the corresponding cuts on surfaces.  

\begin{figure}[t!]
    \centering
    \includegraphics[width=0.45\textwidth]{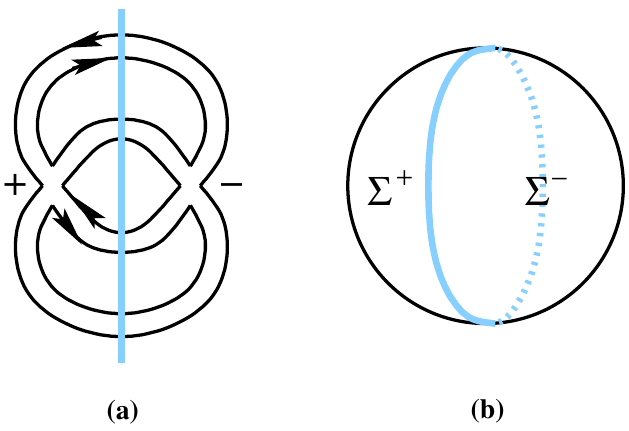}
    \caption{An example of a ribbon diagram with cuts, and its associated surface. \textbf{(a):} This ribbon diagram has a unique extension of the propagator cuts into the plaquettes. \textbf{(b):} The corresponding surface is $\Sigma=S^2$, and the cut decomposes it into two disks $\sigp$, $\sigm$.}
    \label{ffsimcut}
\end{figure}

There are indeed many ribbon diagrams for which this works: An example is shown in Fig.~\ref{ffsimcut}. In such cases, when the cuts across the $G_{+-}$ propagators can be continuously extended across the plaquettes in a unique way, the resulting lines of cuts form a collection of closed circles $S^1$ on $\Sigma$.  Moreover, this collection of $S^1$'s separates $+$ regions and $-$ regions in a way which is globally well-defined for the whole surface.  Thus, cutting $\Sigma$ along this collection of $S^1$'s separates $\Sigma$ into the forward-branch surface $\Sigma^+$ and backward-branch $\Sigma^-$.  The collection of $S^1$'s is then their common boundary, $\p\Sigma^+=\p\Sigma^-$, along which they are glued together to form $\Sigma$.

\begin{figure}[t!]
    \centering
    \includegraphics[width=0.2\textwidth]{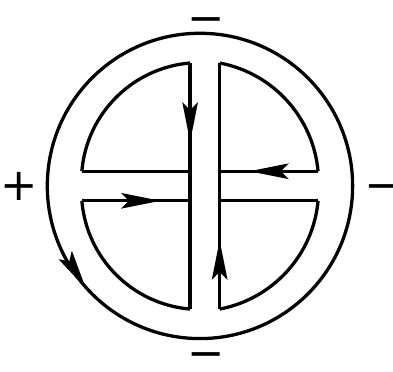}
    \caption{A simple example of a ribbon diagram with an ambiguity in how to connect the cuts across the plaquettes.  There are two plaquettes: The one on the outside has just two adjacent $G_{+-}$ propagators (thus the two cuts can be connected without ambiguity), while the other plaquette has four adjacent $G_{+-}$ edges, giving two inequivalent ways how to connect the four cuts into two nonintersecting lines.}
    \label{ffambigex}
\end{figure}

On the other hand, there are also many ribbon diagrams for which this prescription is incomplete or ambiguous.  An example is shown in Fig.~\ref{ffambigex}.  Upon closer examination, the origin of the ambiguity in this example is clear:  There is a plaquette which has more than two $G_{+-}$ propagators adjacent to it (namely four), and there are two inequivalent ways how the corresponding four cuts can be joined into two nonintersecting lines.  
This makes it clear that the original prescription for extending the cuts across plaquettes to obtain a unique collection of $S^1$'s cuts on $\Sigma$ works precisely for those ribbon diagrams in which each plaquette has at most two $G_{+-}$ propagators adjacent to it.%
\footnote{Of course, in vacuum diagrams considered here, the number of $G_{+-}$ propagators adjacent to any plaquette is always even.}

\begin{figure}[t!]
    \centering
    \includegraphics[width=0.23\textwidth]{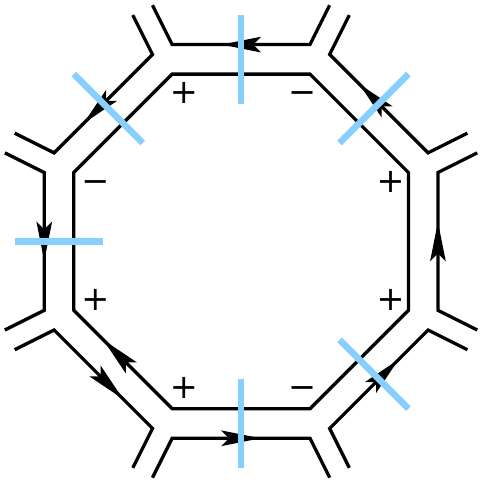}
    \caption{An example of a plaquette with six adjacent $G_{+-}$ propagators, indicating their cuts.}
    \label{ffplcutp}
\end{figure}

\begin{figure}[t!]
    \centering
    \includegraphics[width=0.8\textwidth]{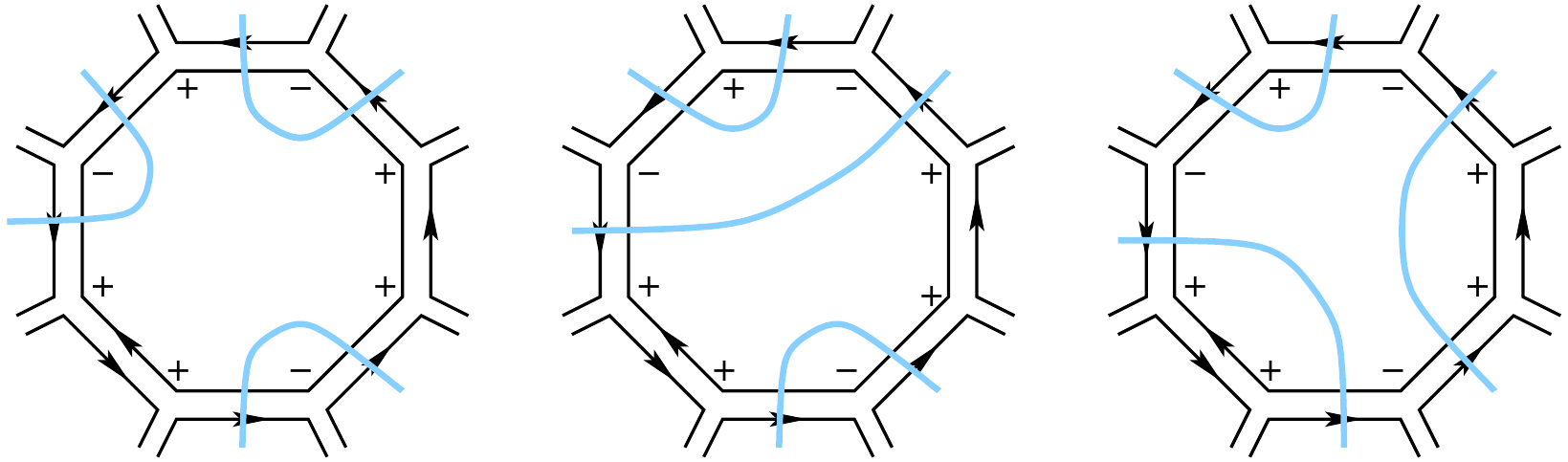}
    \caption{Ambiguities in extending the cuts in Fig.~\ref{ffplcutp} across the plaquette.  In this example, there are five inequivalent ways, of which we show three.}
    \label{ffambig}
\end{figure}

How do we systematically resolve this ambiguity?  Consider a generic plaquette with at least four $G_{+-}$ propagators adjacent to it.  In Fig.~\ref{ffplcutp} we have an example with six.  There is no unique way how to pairwise connect the six cuts illustrated there to form three nonintersecting lines cutting across the plaquette.  In fact, there are five different such pairings, three of which are illustrated in Fig.~\ref{ffambig}.  With the increasing number of adjacent $G_{+-}$ the number of possibilities increases rapidly, and we need a new strategy how to extend the cuts through such plaquettes.  

\begin{figure}[t!]
    \centering
    \includegraphics[width=0.24\textwidth]{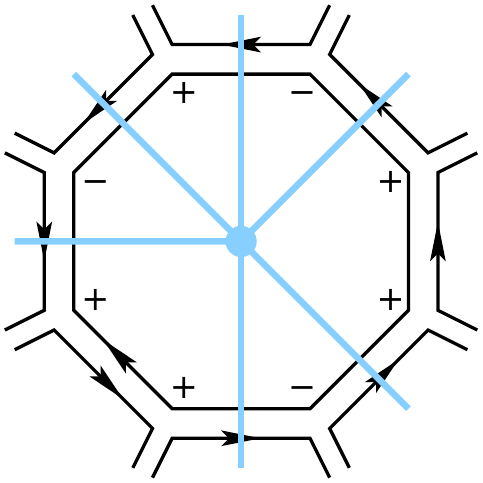}
    \caption{The unique extension of the propagator cuts into the plaquette, by marking the center of the plaquette with a dot and connecting all propagator cuts to the dot.}
    \label{ffpcutg}
\end{figure}

In order to forumulate a unique prescription for extending the cuts, we mark the center of each ambiguous plaquette with a dot, and connect all cuts to the dot in the unique way without forming intersections (see Fig.~\ref{ffpcutg}).  This gives a unique prescription, for any ribbon diagram, how to extend the cuts from the $G_{+-}$ propagators to the full diagram and its associated surface $\Sigma$.  We see that the resulting cut of $\Sigma$ generally does not correspond to a smooth one-dimensional manifold (which would have to be the union of $S^1$'s), but it is described by a graph consisting of a number of dots connected by lines, and drawn on $\Sigma$ in a particular way.%
\footnote{As a general rule, only the centers of those plaquettes which have more than two $G_{+-}$ propagators adjacent to them will be marked with a dot; any plaquette with just two adjacent $G_{+-}$ propagators has an unambiguous cut through it, and no dot is needed in that case.}
Given a diagram $\Delta$, we will denote the graph so constructed by $\Gamma(\Delta)$, and refer to it as the ``graph of cuts'' of the diagram $\Delta$.

\subsection{Topology of worldsheets on the Schwinger-Keldysh time contour}
\label{ssthick}

One could work in this language of cuts given by graphs on worldsheets, but this representation of the cuts is quite cumbersome.  Questions such as: Which graphs are allowed?  How are they mapped to the worldsheet? are not easy to answer in this language.  For example, not every graph, not even every connected graph, is allowed:  It must be bipartite in the sense that it must separate $\Sigma$ into regions that can be consistently labeled alternately by $+$ and $-$.  Moreover, there are way too many allowed graphs, and having to classify them and sum over them would ruin the anticipated simplicity of the topological expansion in string theory.  A much clearer picture emerges when we move away from graphs, and replace them with smooth manifolds.  Indeed, graphs are complicated, but smooth manifolds are simple (at least in low-enough dimensions).

\begin{figure}[t!]
    \centering
    \includegraphics[width=0.24\textwidth]{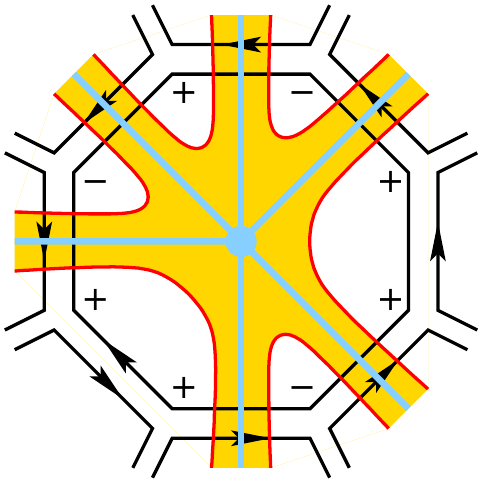}
    \caption{The topologically unique thickening of the graph of cuts $\Gamma$ into a smooth surface with boundaries.  The collection of all such thickenings (denoted here in yellow) across all plaquettes forms the smooth wedge region $\sigw$.}
    \label{ffcutsf}
\end{figure}

How do we associate a graph of cuts $\Gamma$ with a smooth manifold?  Consider one of the ambiguous plaquettes, for example again the one in Fig.~\ref{ffpcutg}.  The graph of cuts across this plaquette is not a smooth manifold, but we can define -- in a topologically unique way -- its ``thickening'' into a smooth two-dimensional surface with smooth one-dimensional boundaries, as indicated in the example of Fig.~\ref{ffcutsf}.  Moreover, these two-dimensional thickenings extend smoothly across all adjacent propagators into neighbouring plaquettes, forming a globally well-defined smooth manifold with non-empty smooth boundary.  We refer to this manifold as the ``wedge region'' of $\Sigma$, and denote it by $\sigw$.%
\footnote{For readers viewing this paper in color, we note note that the wedge regions $\sigw$ are systematically depicted in our Figures in yellow.}
It is this wedge region $\sigw$ that represents the topology of the cuts, connecting $\sigp$ and $\sigm$ into the original smooth surface $\Sigma$.

Thus, we have reached one of the main and perhaps most surprising points of this paper:  The turnaround point $t_\wedge$ on the Schwinger-Keldysh contour, where the forward branch $C_+$ is connected to the backward branch $C_-$, is from the worldsheet point of view topologically two-dimensional!  The cuts connecting the forward and backward parts of $\Sigma$ are not boundaries between $\sigp$ and $\sigm$, but are themselves two-dimensional surfaces $\sigw$.

In the remainder of this Section~\ref{ssnneq}, we will demonstrate in detail that $\sigw$ can have an arbitrarily complicated topology (\ie , any finite number of connected components, handles, and boundaries connecting it to $\sigp$ and $\sigm$), and thefore carries its own genus expansion.

\subsection{The triple decomposition of $\Sigma$}

We have just found that the natural way how to think about the ``cut'' between the forward and backward part of $\Sigma$ is to represent it by a smooth two-manifold with boundaries, not by a one-dimensional graph.  It is this triple decomposition of worldsheets $\Sigma$ into the forward surface $\sigp$, backward surface $\sigm$, and the wedge region $\sigw$ which emerges universally from the large-$N$ expansion.

A simple example is the surface associated with the diagram in Fig.~\ref{ffsimcut}(a), with $\Sigma=S^2$ and the following triple decomposition,
\be
\vcenter{\hbox{\includegraphics[width=1.2in]{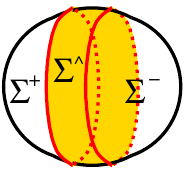}}}\  .
\label{eesfce}
\ee
Our next task is to classify all possible triple decompositions of $\Sigma$ that can emerge from actual ribbon diagrams.  

\begin{figure}[t!]
    \centering
    \includegraphics[width=0.5\textwidth]{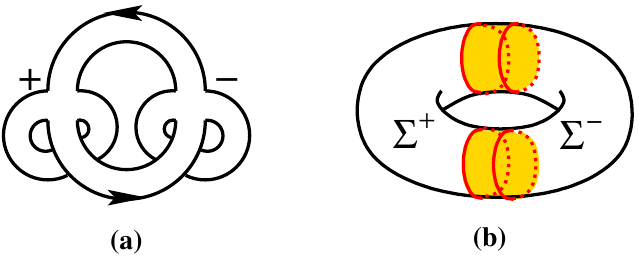}
    \caption{An example yielding more than one connected components of $\Sigma^\wedge$.  \textbf{(a):} A ribbon diagram with $\Sigma=T^2$.  \textbf{(b):} The corresponding triple decomposition of $\Sigma$, with $\Sigma^\pm$ each a cylinder, and $\sigw$ a union of two cylinders.}
\label{ffsk16}
\end{figure}

First of all, it is easy to find examples where $\sigw$ has more than one connected component, but its graph of cuts is still just a collection of circles.  In Fig.~\ref{ffsk16}, the graph of cuts $\Gamma(\Delta)$ has two connected components, each isomorphic to $S^1$, and no vertices.  Thus, $\sigw$ consists of two disconnected cylinders (see Fig.~\ref{ffsk16}(b)).  

\begin{figure}[b!]
    \centering
    \includegraphics[width=0.5\textwidth]{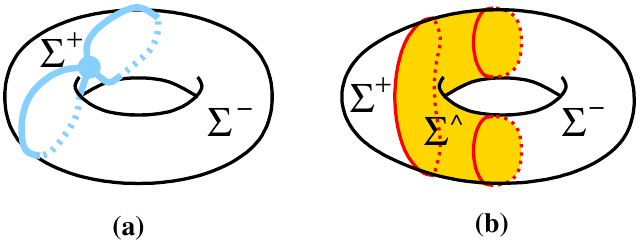}
    \caption{The surface $\Sigma$ that corresponds to the ribbon diagram from Fig.~\ref{ffambigex}, and its triple decomposition.  \textbf{(a):} $\Sigma$ is the torus, $\Sigma^+$ the disk, and $\Sigma^-$ the cylinder.  The cut between them forms a figure-eight graph with one vertex. \textbf{(b):} The triple decomposition of $\Sigma$; the thickening $\sigw$ of the figure-eight graph is the smooth ``pair-of-pants'' surface.}
    \label{ffcutsfv}
\end{figure}

The simplest graph of cuts with at least one vertex is the figure-eight graph.  It can appear in various ribbon diagrams and also be drawn in various inequivalent ways on surfaces.  One example of a ribbon diagram with the figure-eight $\Gamma$ is in Fig.~\ref{ffambigex}, with the associated surface and its triple decomposition depicted in Fig.~\ref{ffcutsfv}.  
Iterating such constructions shows immediately that connected components of $\sigw$ can have an arbitrarily high number of boundary components.

\begin{figure}[t!]
    \centering
    \includegraphics[width=0.18\textwidth]{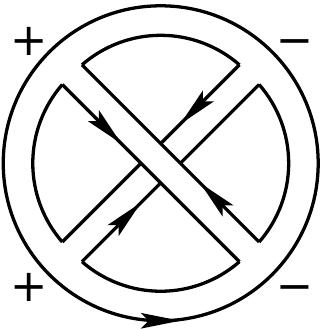}
    \caption{A simple ribbon diagram illustrating that $\Sigma^\wedge$ can be of higher genus.}
    \label{ffwhiteh}
\end{figure}

Next, one wonders about higher genus: Can $\sigw$ with handles also emerge from consistent ribbon diagrams?  To show that the answer is yes, consider the diagram in Fig.~\ref{ffwhiteh}.  
This example gives us an opportunity to introduce a useful mathematical notion, known as the Whitehead reduction of a ribbon diagram:  Given a ribbon diagram $\Delta$ with two distinct vertices of orders $2+k$ and $2+\ell$ connected by a propagator, define the ``Whitehead reduction of $\Delta$'' along this propagator by shrinking the propagator to zero length, thus replacing the two vertices with one composite vertex of order $2+k+\ell$. Since we wish to keep track of the information in the triple decomposition of $\Sigma$, we allow only those Whitehead reductions that do not change this decomposition, \ie, Whitehead reductions along $G_{++}$ propagators and $G_{--}$ propagators are allowed, but Whitehead reductions along the $G_{+-}$ propagators are not.  Two ribbon diagrams that differ by a sequence of allowed Whitehead reductions correspond to the same triple decomposition into $\sigp$, $\sigm$ and $\sigw$.
\begin{figure}[b!]
    \centering
    \includegraphics[width=0.2\textwidth]{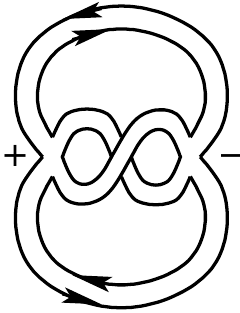}   
    \caption{This diagram is obtained from that in Fig.~\ref{ffwhiteh} by Whitehead reduction, therefore it corresponds to the same surface $\Sigma$ and the same triple decomposition.}
    \label{ffwhat}
\end{figure}

\begin{figure}
    \centering
    \includegraphics[width=0.35\textwidth]{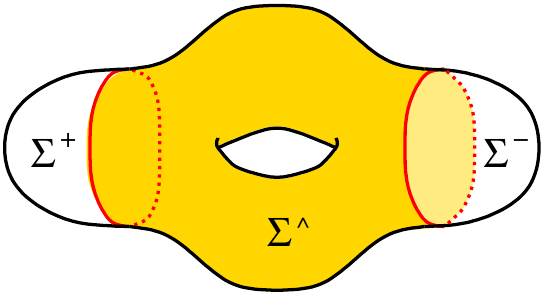}
    \caption{Surface $\Sigma=T^2$ corresponding to the ribbon diagrams in Fig.~\ref{ffwhiteh} and Fig.~\ref{ffwhat}, and its triple decomposition.  While both $\Sigma^+$ and $\Sigma^-$ are disks, $\sigw$ is a surface with two boundaries and a handle.}
    \label{fffav}
\end{figure}

Returning now to our example from Fig.~\ref{ffwhiteh}, we see that the diagram can be simplified by two Whitehead reductions to that depicted in Fig.~\ref{ffwhat}.  Both of these diagrams should thus lead to the same triple decomposition of their underlying surface $\Sigma=T^2$, which we can easily determine by direct inspection:  Since both $\sigp$ and $\sigm$ will be disks, $\sigw$ has to have two boundaries and a handle, as shown in Fig.~\ref{fffav}.
We conclude that the wedge region can indeed carry a handle.  

\subsection{Combinatorial picture of $\Sigma^+$, $\Sigma^-$ and $\Sigma^\wedge$}
\label{sscomb}

In order to prepare the ground for showing that arbitrarily high genera in $\sigw$ can also occur, it will be useful to develop a combinatorial approach to the ribbon diagrams, their associated surfaces $\Sigma$ and their triple decomposition into $\sigp$, $\sigm$ and $\sigw$.

Consider a surface $\Sigma$, obtained from a ribbon diagram $\Delta$.  The ribbon diagram provides a cellular decomposition of $\Sigma$:  The vertices of the ribbon diagram are the zero-dimensional cells, the propagators represent the one-dimensional cells (or edges), and the plaquettes the two-dimensional cells of this cellular decomposition.  For cellular decompositions, the Euler number $\chi(\Sigma)$ of a given surface $\Sigma$ is simply calculated as $\chi(\Sigma)=V-P+L$, with $V$ the number of vertices, $P$ the number of edges, and $L$ the number of plaquettes. We already used this formula in Eqn.~(\ref{eeeuler}), in our review of the large-$N$ expansion in equilibrium.

Now we can use the cellular decomposition of $\Sigma$ implied by $\Delta$ to \textit{define} $\sigp$, $\sigm$ and $\sigw$ by assigning the various element of this cellular decomposition to belong to the three parts of the triple decomposition.  

First, recall that all vertices in $\Delta$ are labeled as either $+$ or $-$, and consequently each propagator is labeled by the two signs indicating the vertices it connects.  We will use the following notation:
\bea
V_+ &=&\textrm{the number of $+$ vertices},\nonumber\\
V_- &=&\textrm{the number of $-$ vertices},\nonumber\\
P_+ &=&\textrm{the number of $G_{++}$ propagators},\nonumber\\
P_- &=&\textrm{the number of $G_{--}$ propagators},\nonumber\\
P_{+-}&=&\textrm{the number of $G_{+-}$ and $G_{-+}$ propagators},\nonumber\\
L_+&=&\textrm{the number of plaquettes (or closed loops) with only $G_{++}$ adjacent propagators},\nonumber\\
L_-&=&\textrm{the number of plaquettes with only $G_{--}$ adjacent propagators},\nonumber\\
L_{+-}&=&\textrm{the number of plaquettes with a non-zero number}\nonumber\\
&&\qquad\qquad\qquad\qquad\qquad\qquad\qquad\qquad\qquad
\textrm{of $G_{+-}$ (or $G_{-+}$) adjacent propagators}.\nonumber
\eea
We now subdivide the elements of the cellular decomposition of $\Sigma$ into those belonging to $\sigp$, $\sigm$ and $\sigw$ as follows:
\begin{itemize}
\item
All $+$ vertices, all $G_{++}$ propagators and all plaquettes with only $G_{++}$ adjacent propagators belong to $\sigp$;
\item
All $-$ vertices, all $G_{--}$ propagators and all plaquettes with only $G_{--}$ adjacent propagators belong to $\sigm$;
\item
All $G_{+-}$ (and $G_{-+}$) propagators and all plaquettes with a non-zero number of $G_{+-}$ (or $G_{-+}$) propagators belong to $\sigw$.
\end{itemize}
This is our combinatorial definition of the triple decomposition of $\Sigma$, in terms of the cellular decomposition defined by the underlying ribbon diagram.

We can now define the ``cellular Euler numbers'' associated with the ingredients of the ribbon diagram $\Delta$ that have been assigned to $\Sigma^\pm$ and $\Sigma^\wedge$ as follows:

\be
\chi_+(\Delta)=V_+-P_++L_+,\qquad
\chi_-(\Delta)=V_--P_-+L_-,
\label{eechipm}
\ee
and
\be
\chi_\wedge(\Delta)=-P_{+-}+L_{+-}.
\label{eechiw}
\ee
It is straightforward to show that the cellular Euler numbers so defined are equivalent to the standard topological definition of the Euler numbers of $\sigp$ and $\sigm$ as topological manifolds with boundaries:
\be
\chi_+(\Delta)=\chi(\Sigma^+),\qquad\chi_-(\Delta)=\chi(\Sigma^-).
\ee
Indeed, this follows from the simple observation that the elements of the cellular decomposition of $\Sigma$ that we assigned to $\sigp$ and $\sigm$ form a cellular decomposition of those surfaces with boundaries, and our definition of $\chi_\pm$ in (\ref{eechipm}) coincides with the standard expression for $\chi(\Sigma^\pm)$ in terms of this cellular decomposition.

It is perhaps a little less immediate to see that the cellular Euler number $\chi_\wedge(\Delta)$ defined in (\ref{eechiw}) is also the topological Euler number of the surface $\sigw$ with boundary whose construction we presented in Section~\ref{ssthick}.  First of all, the elements of the cellular decomposition of $\Sigma$ that we assigned to $\sigw$ do \textit{not} give a cellular decomposition of a surface: There are only edges and plaquettes, but no vertices, and these ingredients do not give a closed submanifold in $\Sigma$.  It is easy, however, to construct an honest cellular decomposition of $\sigw$ by refining the elements that we assigned to $\sigw$. First, add vertices at the ends of all the $G_{+-}$ (and $G_{-+}$) propagators, and think of them as the points at the boundaries between $\sigw$ and $\sigp$ or $\sigm$.  Then connect these vertices by new edges, with each edge simply following these boundaries within each plaquette, as indicated in Fig.~\ref{ffcutsf}.  The addition of these vertices and edges to the ingredients previously assigned to $\sigw$ defines a cellular decomposition of $\sigw$, as a closed manifold with boundary.  Essentially, the new ingredients just add the boundary $S^1$ components to $\sigw$, without changing the alternating sum of the vertices, edges and plaquettes.  We conclude that 
\be
\chi_\wedge(\Delta)=\chi(\Sigma^\wedge).
\ee

Equipped with this combinatorial picture of the triple decomposition, we can now show that $\sigw$ of arbitrarily high genus can indeed emerge from ribbon diagrams.

\begin{figure}[t!]
    \centering
    \includegraphics[width=0.6\textwidth]{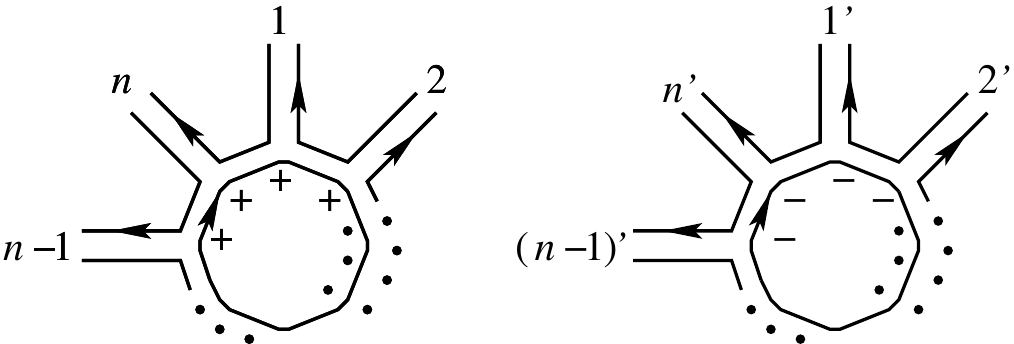}
    \caption{Construction of the ribbon diagram whose $\Sigma^\wedge$ is a higher-genus surface with two boundary components, depicted in Figure~\ref{ffhighsf}. Prepare two ribbon diagrams with $n$ loose ends each as indicated, and connect pairwise the ends labeled by $i$ and $i'$: $1$ to $1'$, $2$ to $2'$, $\ldots$, $n$ to $n'$.  Note that with this order of gluing the ends, the resulting ribbon diagram will have only the total of 3 plaquettes if $n$ is odd, or 4 plaquettes if $n$ is even.}
    \label{ffhigh}
\end{figure}

Consider a ribbon diagram, constructed from ingredients shown in Fig.~\ref{ffhigh}:  Two ribbon diagrams with $n$ loose ends.  If we glue the end marked 1 with $n'$, 2 with $n'-1$, $\ldots$ and $n$ with 1${}'$, the surface $\Sigma$ associated with the resulting diagram is the sphere. Indeed, in this case we have
\be
V_+=V_-=P_+=P_-=P_{+-}=n, \quad L_+=L_-=1,
\label{eenexa}
\ee
and $L_{+-}=n$, implying that $\chi(\sigp)=\chi(\sigm)=1$, $\chi(\sigw)=0$, and $\chi(\Sigma)=2$.  The triple decomposition is the one we found in (\ref{eesfce}): $\sigp$ and $\sigm$ are disks, and $\sigw$ is a cylinder. 

\begin{figure}[b!]
    \centering
    \includegraphics[width=0.35\textwidth]{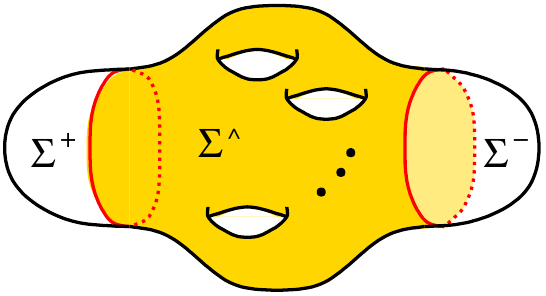}
    \caption{The worldsheet topology whose $\sigw$ has $h$ handles and two boundary components, and $\sigp$, $\sigm$ are both disks.  This surface is obtained from the ribbon diagram construction depicted in Figure~\ref{ffhigh}, with $n=2h+1$ or $n=2h+2$.}
    \label{ffhighsf}
\end{figure}

On the other hand, if we glue the loose ends in the order indicated in Fig.~\ref{ffhigh}, we obtain a surface whose cellular decomposition is characterized by the same numbers as in (\ref{eenexa}), while the number $L_{+-}$ of ${+-}$ plaquettes changes from $n$ to just 1 if $n$ is odd, and to 2 if $n$ is even.  Thus, for odd $n=2h+1$ or even $n=2h+2$, we see that $\chi(\sigw)=-2h$.  Since $\Sigma^\pm$ are disks, $\sigw$ has two boundary components. We conclude that $\sigw$ resulting from the construction in Fig.~\ref{ffhigh} is the surface with two boundaries and $h$ handles (see Fig.~\ref{ffhighsf}).  This demonstrates that wedge regions $\sigw$ obtained from ribbon diagrams can have connected components with an arbitrarily high number of handles.

\subsection{Measure once, cut twice}

Our analysis of the large-$N$ expansion of non-equilibrium systems revealed one, perhaps surprising, fact:  The time instant $t_\wedge$ where the forward branch of the Schwinger-Keldysh time contour meets the backward branch does not cut the worldsheet $\Sigma$ into its forward and backward parts $\sigp$ and $\sigw$ connected along a common one-dimensional boundary -- instead, the worldsheet region $\sigw$ corresponding to $t_\wedge$ is topologically two-dimensional, and even carries its own genus expansion.

Having seen that the wedge region $\sigw$ can have components with arbitrary numbers of boundaries and high genus, there are some natural follow-up questions about $\sigw$ and how it is connected to the forward and backward regions $\sigp$ and $\sigw$.  

Does $\sigw$ always have to have non-empty boundaries with both $\sigp$ and $\sigm$?  The answer is yes, in the following sense:  There are certainly ribbon diagrams, such that their associated surface is $\Sigma=\sigp$ or $\Sigma=\sigm$, and $\sigw$ is empty.  But if $\sigw$ is non-empty, it has to have a non-empty boundary both with $\sigp$ and with $\sigm$.  The proof is simple:  In the combinatorial description, $\sigw$ is built from lines representing $G_{+-}$ propagators, and the $+-$ plaquettes.  If $\sigw$ is non-empty, it contains at least one $G_{+-}$ propagator.  This propagator has to have a place to end, at both ends.  On the $+$ side, the propagator can only end at a non-empty boundary with $\sigp$; similarly, on the $-$ side, it must end on a non-empty boundary with $\sigm$.

We can summarize this structure in a simple slogan: \textit{Measure once, cut twice!} If you find that $\Sigma$ has both $\sigp$ and $\sigm$ non-empty, you must cut; and if you cut, you must cut twice.  The first ``cut'' indicates the location within $\Sigma$ of the boundary between $\sigp$ and $\sigw$ (which we denote by $\p_+\Sigma$), and the second ``cut'' indicates the location of the boundary between $\sigw$ and $\sigm$ (which we denote by $\p_-\Sigma$).  Of course, $\p_+\Sigma$ is just a collection of $n$ circles and $\p_-\Sigma$ is a collection of $n'$ circles; note that $n$ does not have to equal $n'$.

It is intriguing to find that the structure of worldsheet ``cuts'' is so much richer in comparison to the simple propagator cuts known from standard quantum field theory of particle physics.

\subsection{Dual picture}
\label{ssdual}

Ribbon diagrams exhibit a very useful duality property, closely related to what mathematicians call Poincar\'e duality in topology of manifolds.  Each ribbon diagram $\Delta$ defines uniquely another,  dual ribbon diagram $\Delta^\star$, as follows:  Each plaquette of $\Delta$ is associated with a vertex in $\Delta^\star$. Whenever two plaquettes in $\Delta$ share an edge, the corresponding vertices in $\Delta^\star$ are connected by a ribbon.  All ribbons attached to a given vertex in $\Delta^\star$ in the same cyclic order as the order of their dual edges around the original plaquette in $\Delta$.  As a result, each plaquette in $\Delta^\star$ is associated with a unique vertex in $\Delta$.  It is easy to see that with this construction, $\Sigma(\Delta^\star)=\Sigma(\Delta)$ (and consequently $\chi(\Sigma(\Delta^\star))=\chi(\Sigma(\Delta))$), and $(\Delta^\star)^\star=\Delta$.  In particular, the cellular decompositions of $\Sigma$ provided by a diagram $\Delta$ and its dual $\Delta^\star$ are dual to each other in the sense of cellular decompositions.

We can use this duality to shed more light on the $\sigw$ region.  In the combinatorial description of $\Sigma$ using a ribbon diagram $\Delta$, it was perhaps surprising that we assigned only propagators and plaquettes to $\sigw$, \ie, one-dimensional and two-dimensional cells, but no vertices.  The dual picture using $\Delta^\star$ reveals why that was so: The plaquettes of $\Delta$ correspond to vertices in the dual picture, and the $G_{+-}$ propagators that traverse across $\sigw$ from the $\p_+\Sigma$ boundary to the $\p_-\Sigma$ boundary turn in the dual picture to lines connecting those vertices.  Thus, dualizing the formula (\ref{eechiw}) for the Euler number of $\sigw$, we see that contributions to $\chi(\sigw)$ come only from vertices and propagators of $\Delta^\star$, so only zero- and one-dimensional components contribute.  These components of $\Delta^\star$ of course form nothing other than the graph of cuts $\Gamma(\Delta)$.  In this sense, the topological information about $\sigw$ can be encoded in cellular data not involving cells of dimension two.  

\begin{figure}[t!]
    \centering
    \includegraphics[width=0.4\textwidth]{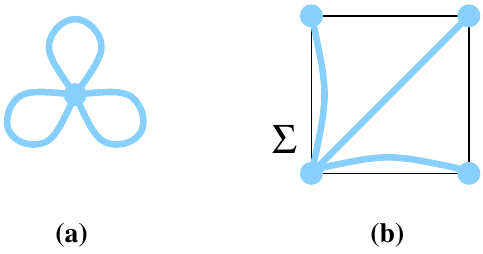}
    \caption{Surfaces with distinct $\sigw$'s but with the same graph of cuts $\Gamma$. In this example, $\Gamma$ is the trefoil graph. \textbf{(a):}  Here $\Gamma$ is drawn on the sphere, and $\sigw$ is the sphere with four boundaries.  \textbf{(b):} Here $\Sigma$ is the torus, represented as a square with the opposite sides pairwise identified.  The indicated graph of cuts is again the trefoil. The triple decomposition of this surface reproduces that of Fig.~\ref{fffav}.}
    \label{ffsk15}
\end{figure}

This does not mean that we should abandon our smooth-surface representation of $\sigw$ and revert back to the graph description:  The classification of $\sigw$ as surfaces with smooth boundaries is much more transparent than the classification of the corresponding graphs and the ways how they can be drawn on surfaces.  In particular, without keeping track of how the graph of cuts $\Gamma$ is drawn on $\Sigma$, the graph itself does not contain enough information to reconstruct the topology of $\sigw$.  Take for example the trefoil graph, depicted in Fig.~\ref{ffsk15}(a).  This graph can be drawn on the sphere, in a topologically unique way.  This configuration indeed corresponds to a particular ribbon diagram, whose $\Sigma=S^2$ and $\sigw$ is the sphere with four boudaries.  The trefoil graph can also be drawn on a torus, in several inequivalent ways.  First, if drawn in a local patch of the torus, it again gives the same $\sigw$ as on the sphere.  Or it can be drawn such that all three cycles of the trefoil are noncontractible and mutually homotopically inequivalent, as in Fig~\ref{ffsk15}(b).  This describes the configuration in Fig.~\ref{ffwhat}, and $\sigw$ is the torus with two boundaries.  Of course, both of these $\sigw$ topologies (as well as the graph $\Gamma$ itself) have the same Euler number, $\chi=-2$.  

Indeed, this ambiguity is not at all surprising -- in order to keep track of how $\Gamma$ is drawn on $\Sigma$, we have just learned that it is natural to interpret it as a ribbon subdiagram of the dual ribbon diagram $\Delta^\star$.  As a two-dimensional surface with boundaries, this ribbon diagram corresponding to $\Gamma$ is indeed just the thickening of $\Gamma$ into $\sigw$ that we introduced in Section~\ref{ssthick}.  

\subsection{Grothendieck's \textit{dessins d'enfants} make an appearance}

Over its lifetime, string theory has demonstrated an extraordinary ability to make meaningful connections to many diverse areas of modern mathematics.  These connections have been very fruitful both for mathematics and physics.  In this subsection, we take a brief detour to point out one unexpected connection between the structure of the non-equilibrium string diagrams and objects that have been studied extensively in pure mathematics, under the name of Grothendieck's \textit{dessigns d'enfants}.  Tho readers interested in the physical picture of non-equilibrium string perturbation theory should feel free to skip this subsection and go directly to Section~\ref{ssnspt}, where our main results are stated.

Since the notion of \textit{dessins d'enfants} (see, \eg , \cite{leila,graphs,guillot,dessins}) was first introduced by Grotendieck in his 1984 \textit{Esquisse d'un Programme}, \textit{dessins} have been found to relate remarkably many diverse areas in pure mathematics (including such arkane concepts as the absolute Galois group%
\footnote{The absolute Galois group $\textrm{Gal}(\overline{\mathbb Q}/{\mathbb Q})$ is defined as the group of authomorphisms of the algebraic numbers $\overline{\mathbb Q}$ which fix the rational numbers ${\mathbb Q}$ \cite{galois}.}
and its faithful action on various categories \cite{dessins}), and it is only fitting that they should appear in string theory.%
\footnote{In an unrelatex context, \textit{dessins d'enfants} also appeared previously in string theory in certain brane engineering constructions \cite{dessbe} and Calabi-Yau compactifications \cite{desscy,desscyy}.}
For our purposes, a \textit{dessin d'enfant} can be defined as a connected graph, consisting of a finite number of vertices and lines, and drawn on a two-dimensional surface $\Sigma'$ such that no two vertices coincide on $\Sigma'$ and no two lines intersect on $\Sigma'$, and such that two additional conditions are satisfied: (i) the graph is bipartite in the following sense: each vertex is labeled either black or white, with each line of the graph connecting a black vertex with a white one; and (ii) the complement of the graph in $\Sigma'$ is topologically a collection of disks.  

Now we can show that there is a close relation bewteen \textit{dessins d'enfants} and the wedge-region part of our ribbon graphs.  Imagine asking the following question:  How do we keep track of only that part of the ribbon diagram that defines the $\sigw$ region of its associated surface?  This question is answered as follows.  Consider a ribbon diagram $\Delta$, and draw it on its associated surface $\Sigma$ with triple decomposition $\sigp$, $\sigm$ and $\sigw$.  Erase the $\sigp$ and $\sigw$ parts of $\Sigma$, keeping only the wedge region $\sigw$.  Glue in a disk inside each of the boundary components of $\sigw$, place a new vertex in the center of each such disk, and label it $+$ or $-$ depending on whether the disk replaces a boundary with $\sigp$ or $\sigm$.  All ribbon propagators are now ending on the edges of the glued-in disks; extend them to the vertex at the center of the disk, without intersections.  This defines a new, ``reduced'' ribbon diagram, whose associated surface is $\sigw$ with the boundary components filled in with the disks.  All detailed information about how the original ribbon diagram extends into the $\sigp$ and $\sigm$ regions has now been erased, so the resulting reduced ribbon diagram encodes only the information about $\sigw$.  

We now observe that each such reduced ribbon diagram defines a unique \textit{dessin d'enfant}:  Label each $+$ vertex as black, and each $-$ vertex as white, and note that all the axioms of \textit{dessins} are satisfied by our reduced ribbon diagram.  In turn, every \textit{dessin d'enfant} is realized by at least one ribbon diagram in this way.  More precisely, we can define an equivalence relation on the original ribbon diagrams, by declaring two ribbon diagrams equivalent if they may differ only in their $\sigp$ and $\sigm$ regions, but give the same reduced ribbon diagram when our procedure is followed.  Two ribbon diagrams correspond to the same \textit{dessin d'enfant} if they belong to the same equivalence class.  For example, the ribbon diagrams in Figs.~\ref{ffwhiteh} and \ref{ffwhat} are in the same equivalence class, and the \textit{dessin d'enfant} corresponding to them can be drawn like this:
\be
\vcenter{\hbox{\includegraphics[width=1.2in]{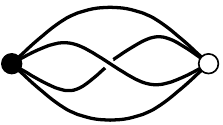}}}\ \ \ .
\nonumber
\ee
This \textit{dessin} is supposed to be visualized as being drawn on a torus, and the $\sigw$ that corresponds to this \textit{dessin} is depicted in Fig.~\ref{fffav}.

In turn, two ribbon diagrams from distinct equivalence classes correspond to distinct \textit{dessins}.  We conclude that there is a one-to-one correspondence between \textit{dessins d'enfants} and the equivalence classes of all ribbon diagrams defined above, represented by the reduced ribbon diagrams.

We do not have any immediate use in non-equilibrium physics for this connection to \textit{dessins d'enfants}, yet we find it fascinating that they do naturally appear in the structure of non-equilibrium string perturbation theory, and are related so intimately to the most interesting portion $\sigw$ of the worldsheet, associated with the crossing from the forward to the backward branch of the Schwinger-Keldysh time contour.  

\subsection{Non-equilibrium string perturbation theory}
\label{ssnspt}

After this thorough analysis of the surfaces $\Sigma$ that can emerge from ribbon diagrams in our large-$N$ theory of matrix degrees of freedom out of equilibrium, we are ready to formulate the main lessons about the dual string theory expansion.  Which surfaces contribute to the expansion?  If we make no additional assumptions about the dynamics of the large-$N$ system, \ie , assume no ``hidden identities'' of individual ribbon diagrams (or among groups of ribbon diagrams) that would make some contributions vanish, then as we have seen above, all possible ttriple decompositions of worldsheets result from consistent ribbon diagrams.%
\footnote{For specific systems, there might be additional identities that make some classes of surfaces drop out from the sum; those can be studied on a case-by-case basis.  Here we concentrate on the universal predictions about non-equilibrium string perturbation theory, following solely from the topology of the large-$N$ expansion, without any additional dynamical assumptions.}

In non-equilibrium string perturbation theory, the partition function is expressed as a refined topological expansion over worldsheet surfaces, 
\begin{figure}[t!]
    \centering
    \includegraphics[width=0.4\textwidth]{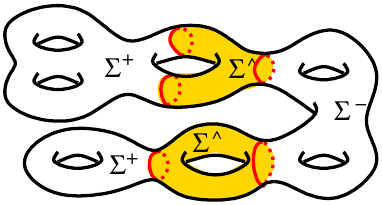}
    \caption{A typical string topology contributing to the non-equilibrium string perturbation theory.}
\end{figure}
\be
\CZ=\sum_{h=0}^\infty\left(\frac{1}{N}\right)^{2h-2}\sum_{\substack{\textrm{triple decompositions}\\ \chi_++\chi_-+\chi_\wedge=2-2h}}\CF_{\sigp,\sigm,\sigw}(\lambda,\ldots),
\label{eeneqspt}
\ee
This formula is the central result of this paper: In non-equilibrium string theory, the genus expansion into a sum over connected surfaces $\Sigma$ known from equilibrium is further refined into a sum over triple decomposition of each surface $\Sigma$ into its forward part $\sigp$, backward part $\sigm$ and the wedge part $\sigw$ which corresponds to the time instant $t_\wedge$ where the two branches $C_+$ and $C_-$ of the Schwinger-Keldysh time contour meet.  We stated the result here for the partition function $\CZ$, but the same expansion is expected of correlation functions of local observables as well.  

In our derivation of this result from the original large-$N$ system, the individual contribution $\CF_{\sigp,\sigm,\sigw}$ of each triple decomposition is weighted by the power of $N$ given by the total Euler number $\chi(\Sigma)=2-2h$.  Thus, the term at a fixed order $h$ in the string coupling is further refined into a sum over all triple decompositions of $\Sigma$ with that genus $h$ into $\sigp$, $\sigm$ and $\sigw$, subject only to the condition that $\Sigma$ be connected.  At this stage, individual triple decompositions are still weighted just by the overall Euler number $\chi(\Sigma)$, with $1/N$ the only parameter of the expansion.  This may be the limit of how far the large-$N$ expansion arguments can take us, in predicting the universal properties of the dual string theory.

However, once we identify $1/N$ with the string coupling constant $g_s$, one can use our experience with critical string theory at equilibrium to speculate that a more refined weighting should be possible.  For example, one can imagine dialing different values of the string coupling on the forward and backward branches of the time contour (let's call them $g_+$ and $g_-$), or a different value of the string coupling in the asymptotic future at $t_\wedge$ (which we naturally call $g_\wedge$).  Indeed, in critical string theory in equilibrium, there are many examples where the string coupling ``constant'' -- being given by the vacuum expectation value of the dilaton field $\Phi$ as $g_s=\langle e^\Phi\rangle$ -- is dependent on the spacetime location, no longer necessarily equal to a fixed value set by $1/N$.  Assuming that on the Schwinger-Keldysh contour $g_s$ can take such three different values $g_\pm$ and $g_\wedge$ in its three different regions, each term in the perturbation theory sum (\ref{eeneqspt}) would then be weighted by the more refined weight
\be
g_+^{-\chi(\sigp)}g_-^{-\chi(\sigm)}g_\wedge^{-\chi(\sigw)},
\ee
replacing the overall $g_s^{-\chi(\Sigma)}$ that we obtained from the $1/N$ expansion.  In order to see whether such a possibility is realized, we would need to know more about the worldsheet dynamics of strings away from equilibrium.

\begin{figure}[t!]
    \centering
    \includegraphics[width=0.25\textwidth]{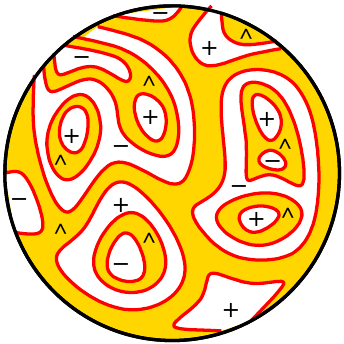}
    \caption{Illustration of the possible infinite proliferation of triple decompositions for a given $\Sigma$ (here illustrated for $\Sigma=S^2$), if connected components of $\sigp$ and $\sigm$ with non-negative Euler number (and no additional insertions of observables inside them) are not identically zero.}
    \label{ffsk13}
\end{figure}

Having shown that all topologies can appear in the triple decompositions of worldsheet surfaces, one can reorganize the question and ask, given a surface $\Sigma$, for a full classification of all its possible triple decompositions.  Such decompositions are fully classified in terms of the discrete topological data about $\sigp$, $\sigm$ and $\sigw$:  Their numbers of handles, and numbers of boundary components.  However, without making any additional assumptions about the worldsheet dynamics, the number of distinct triple decompositions of a fixed surface $\Sigma$ is infinite.  This proliferation of decompositions is illustrated for the sphere in Fig.~\ref{ffsk13}.  We find an infinite number of decompositions of the sphere, with connected components of $\Sigma^\wedge=\Sigma_{0,b}$ given by spheres with $b$ boundaries.  Upon closer inspection, we find that the origin of this proliferation is in the existence of components in $\sigp$, $\sigm$ and $\sigw$ whose Euler number $\chi$ is non-negative:  Indeed, since the Euler number of $\Sigma$ is the sum of the Euler numbers of its triple decomposition, if only those components that have negative $\chi$ were allowed, there would only be a finite number of possible decompositions.

In vacuum diagrams, the components of $\sigp$ and $\sigw$ with non-negative Euler numbers
are disks and cylinders, while in $\sigw$ it is only the cylinder.  When we generalize from vacuum diagrams to  correlation functions of local observables on $\Sigma$, each insertion counts as a ``puncture'' in $\Sigma$, and contributes an additional $-1$ to the Euler number.  In this case, the additional components of $\sigp$ and $\sigw$ causing proliferations in the triple decompositions of $\Sigma$ are also disks with one puncture.  
How to deal with such proliferations?  There are two main options: (a) allow them to be non-zero and perhaps resum the contributions with non-negative $\chi$ to define ``renormalized'' triple decompositions of $\Sigma$, or (b) make an additional assumption about the worldsheet dynamics, declaring that contributions of components of $\sigp$ and $\sigm$ with $\chi\geq 0$ vanish identically.

While Option~(a) might be necessary in some circumstances far from equilibrium, Option~(b) is something we are familiar with from critical string theory in equilibrium.  In critical string theory, string worldsheets inherit a complex structure from the dynamics of worldsheet gravity and its symmetries.  The $\CF_h$ contributions at fixed genus $h$ are given as integrals over moduli spaces of such complex structures.  When the worldsheet is a sphere with fewer than three punctures, such contributions vanish identially, since they are suppressed by the infinite volume of a residual worldsheet gauge symmetry.  In the language of mathematics, only ``stable nodal Riemann surfaces'' \cite{stable} (\ie , surfaces with punctures and with non-negative Euler numbers) contribute to the amplitudes.  This suggests a realization of our Option~(b): In theories where the worldsheet dynamics implies additional worldsheet structure (such as the complex structure), one could propose that the boundaries $\p_+\Sigma$ and $\p_-\Sigma$ in the triple decomposition should be interpreted geometrically as nodes in the Riemann surface, and expect that the components of $\sigp$ and $\sigm$ which carry non-negative Euler numbers vanish identically, in analogy with critical string theory in equilibrium.  

Note that in Option~(b), in order to get a finite sum over triple decompositions, it would not be sufficient to assume that just the components of $\sigp$ and $\sigm$ with strictly positive $\chi$ (\ie , the disks) vanish identically:  There would still be an infinite number of triple decompositions of vacuum diagrams at each order in $g_s$, starting at genus one.

In fact, the list of topological invariants associated universally with our triple decompositions of $\Sigma$ is even richer than just $\chi(\sigp)$, $\chi(\sigm)$ and $\chi(\sigw)$.  We can define $b_+$ to be the number of boundary components in the boundary $\p_+\Sigma$ between $\sigp$ and $\sigw$, and similarly $b_-$ as the number of components in the boundary $\p_-\Sigma$ between $\sigm$ and $\sigw$.  These $b_\pm$ are of course topological invariants, and if we introduce ``fugacities'' $f_+$ and $f_-$ for them, we can weigh each triple decomposition of $\Sigma$ by an additional factor of
\be
f_+{}^{b_+}f_-{}^{b_-}.
\ee

Another set of useful invariants are the numbers of connected components in $\sigp$, $\sigm$ and $\sigw$, which we denote by $n_+$, $n_-$ and $n_\wedge$.  Even if Option~(a) applies, and the disk and cylinder components of $\sigp$ and $\sigm$ turn out not to be zero, there is one way how to reduce the sum over triple decompositions at each genus $h$ to a finite sum:  If we allow only connected $\sigp$ and $\sigm$ to contribute.  This can be arranged by introducing  ``fugacity'' parameters $\gamma_+$ and $\gamma_-$ for the numbers of components $n_+$ and $n_-$, to weigh the contribution of a given triple decomposition by
\be
\gamma_+^{n_+-1}\gamma_-^{n_--1}.
\ee
Presumably, we choose $\gamma_\pm$ to be smaller than one, so that they suppress contributions from higher numbers of connected components of $\Sigma^\pm$.  Sending $\gamma_\pm\to 0$ then keeps only the contributions from connected $\Sigma^\pm$.

The question of whether or not the appropriate refined expansion parameters such as $g_\pm$ and $g_\wedge$, $\gamma_\pm$ or $f_\pm$ do naturally appear in a given string theory is likely to depend on the specific examples, and their string dynamics.  Since in this paper we are only focusing on the universal properties independent of any knowledge about the worldsheet dynamics, such questions are outside of the scope of this paper.  Our universal arguments only reveal the universal existence of the topological invariants to which such hypothetical dynamical expansion parameters could be sensitive.

\section{Strings on the Kadanoff-Baym time contour}

Our analysis of the Schwinger-Keldysh time contour has several straightforward generalizations.  In this section, we present the large $N$ expansion of theories on another popular time contour, relevant particularly for systems at finite temperature $T$ at or near equilibrium, known as the Kadanoff-Baym contour%
\footnote{Sometimes this contour is referred to as the Konstantinov-Perel' contour \cite{svl}.}
\cite{kadb,das,dast,spicka1,spicka2,spicka3,spicka}.  Since the logic of this analysis is a straighforward generalization of our discussion in Section~\ref{ssnneq}, we will be relatively brief.

\subsection{The Kadanoff-Baym contour and finite temperature}

\begin{figure}[t!]
    \centering
    \includegraphics[width=0.45\textwidth]{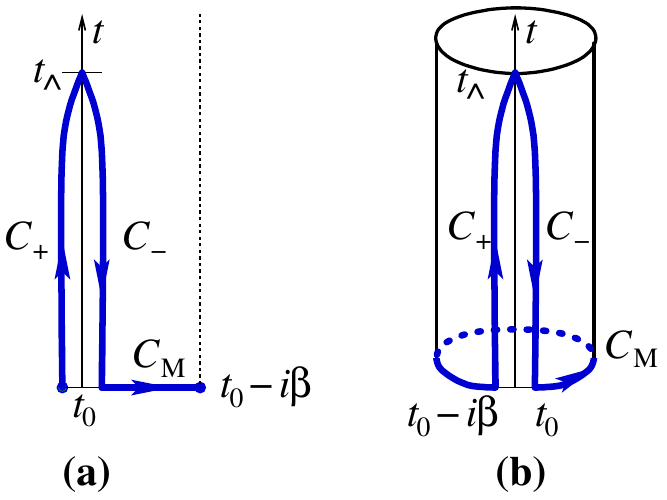}
    \caption{\textbf{(a):} The Kadanoff-Baym time contour $\SC_\beta=C_+\cup C_-\cup C_\rmm$ in the plane of complexified time, with the dashed line indicating the periodicity of observables by $\beta$ in the imaginary time direction. \textbf{(b):} The KMS periodicity properties suggest that the complexified time can be naturally thought of as a cylinder, on which the KB contour is a closed contour with winding number one.}
    \label{ffkbc}
\end{figure}

The Kadanoff-Baym (KB) contour $\SC_\beta$ consists of three segments (see Fig~\ref{ffkbc}(a)):  Besides the forward and backward branches $C_+$ and $C_-$ known from the Schwinger-Keldysh contour, there is a third segment $C_\rmm$ representing an excursion into the imaginary direction by the amount $-i\beta$.  This last segment of the KB contour is referred to as the ``Matsubara'' segment of the KB contour. Indeed, this Matsubara segment would constitute the entire time contour in the standard imaginary-time approach to equilibrium systems at finite temperature known as the Matsubara formalism.  Keeping both the imaginary-time segment and the real-time segments of the KB contour allows us to combine the benefits of the imaginary-time Matsubara formalism with the possibility of studying real-time phenomena at finite temperature.  The condition of thermal equilibrium translates into the so-called Kubo-Martin-Schwinger (KMS) conditions on correlation functions of meaningful quantities.  As a consequence of the KMS conditions, the correlation functions are periodic (or antiperiodic) along the imaginary direction of the complexified time; it is therefore natural to think of the KB contour as a closed contour on the cylinder (see Fig~\ref{ffkbc}(b)).

Now the fields are tripled: $M$ on the $\SC_\beta$ contour can be represented by two fields $M_\pm(t)$ that depend on real time, and one new field $M_\rmm(\tau)$ which depends on the coordinate $\tau$ defined as $\tau=-\im\,t$ along the Matsubara segment $C_\rmm$:
\be
M_\rmm{}^a{}_b(\tau)\equiv M^a{}_b(-i\tau).
\ee
With this definition, $\tau\in[0,\beta)$.  

This triplication of fields means that we have nine \textit{a priori} distinct propagators, defined using the time ordering $\T_{\SC_\beta}$ along the KB contour.  They are denoted by ribbons as in (\ref{eeproppp}-\ref{eepropmm}), but now labeled with three possible indices $+,-,\rmm$ at each end.  The propagators involving the $M_\pm$ fields are as in (\ref{eetimepp}-\ref{eetimemm}).  Then there are four propagators connecting one $M_\rmm$ with either $M_+$ or $M_-$; these are expressed in terms of the Green's functions $G^\lceil$ and $G^\rceil$ known in the non-equilibrium literature as $G$-left and $G$-right \cite{svl},
\bea
\left\langle M_\rmm(\tau)M_\pm(t)\right\rangle&=&g^2\,G^\lceil(\tau,t),\\
\left\langle M_\pm(t)M_\rmm(\tau)\right\rangle&=&g^2\,G^\rceil(t,\tau).
\eea
Finally, we have the $G_{\rmm\rmm}$ propagator, familiar from the Matsubara formalism, and given by the two-point function of $M_\rmm$ along the Matsubara segment.  For clarity, we will again use a uniform two-index notation for all nine propagators in the rest of this section, with the indices running over $+,-,\rmm$.

The vertices are the same as in (\ref{eevert+}-\ref{eevert--}), except now they are labeled by one of the three indices $+,-,\rmm$.

\subsection{Seven-fold decomposition of $\Sigma$}

In understanding the decomposition of $\Sigma$ for the KB contour, we will use the same combinatorial approach that worked for us in Section~\ref{sscomb}.

All ribbon diagrams now have vertices labeled by $+, -$ and $\rmm$.  Consider such a diagram $\Delta$.  It defines a cellular decomposition of its associated surface $\Sigma$.  We wish to construct the decomposition of $\Sigma$ on the KB contour, analogous to the triple decomposition of $\Sigma$ that we found on the Schwinger-Keldysh contour.  We begin by constructing the forward region $\sigp$:  Combinatorially, we define $\sigp$ to be the region whose cellular decomposition consists of all $+$ vertices in $\Delta$, all the $G_{++}$ propagators, and all the plaquettes whose all adjacent propagators are $G_{++}$.  This collection of data indeed defines a cellular decomposition of a surface with boundaries, which will be our $\sigp$.  Repeating the same with $-$ vertices, propagators and plaquettes defines the backward region $\sigm$.  Finally, repeating the same with $\rmm$ vertices, propagators and plaquettes defines $\Sigma^\rmm$, the ``Matsubara region'' of $\Sigma$.

In complete analogy with Section~\ref{sscomb}, we introduce the following notation:
\bea
V_+ &=&\textrm{the number of vertices labeled by $+$},\nonumber\\
P_+ &=&\textrm{the number of $G_{++}$ propagators},\nonumber\\
L_+ &=&\textrm{the number of plaquettes with all their vertices labeled by $+$},\nonumber
\eea
(with similar definitions for $V_-$, $P_-$, $L_-$, $V_\rmm$, $P_\rmm$ and $L_\rmm$).  We 
define the combinatorial Euler numbers $\chi_\pm$ and $\chi_\rmm$, and argue that they are equal to the topological Euler numbers of surfaces with boundaries $\sigp$, $\sigm$ and $\Sigma^\rmm$:
\bea
\chi_+(\Delta)&\equiv& V_+-P_++L_+=\chi(\sigp),\nonumber\\
\chi_-(\Delta)&\equiv& V_--P_-+L_-=\chi(\sigm),\nonumber\\
\chi_\rmm(\Delta)&\equiv& V_\rmm-P_\rmm+L_\rmm=\chi(\Sigma^\rmm).\nonumber
\eea
Next we try to repeat our definition of $\sigw$, and define the regions of $\Sigma$ that correspond to the parts of the KB contour where two of the regions $\sigp$, $\sigm$ or $\Sigma^\rmm$ connect.  First, we define region $\Sigma^{+-}$ by assigning to it all $G_{+-}$ and $G_{-+}$ propagators in $\Delta$, and all the plaquettes with at least one adjacent $G_{+-}$ or $G_{-+}$ propagator but no adjacent $\rmm$ vertices.  We denote by $V_{+-}$, $P_{+-}$ and $L_{+-}$ the numbers of vertices, propagators and plaquettes so assigned to $\Sigma^{+-}$.  Next, we similarly define regions $\Sigma^{+\rmm}$ and $\Sigma^{-\rmm}$ by repeating the same steps which defined $\Sigma^{+-}$.  

Precisely as in the case of $\sigw$ in Section~\ref{sscomb}, these combinatorial data contain no vertices, and they do not define a cellular decomposition of the three surfaces. We can still define the cellular Euler numbers 
\bea
\chi_{+-}(\Delta)&\equiv& V_{+-}-P_{+-}+L_{+-},\nonumber\\
\chi_{+\rmm}(\Delta)&\equiv& V_{+\rmm}-P_{+\rmm}+L_{+\rmm},\nonumber\\
\chi_{-\rmm}(\Delta)&\equiv& V_{-\rmm}-P_{-\rmm}+L_{-\rmm},\nonumber
\eea
and ask whether they are equal to the topological Euler numbers of the surfaces $\Sigma^{+-}$, $\Sigma^{+\rmm}$ and $\Sigma^{-\rmm}$.  In contrast to the Schwinger-Keldysh case, here we find that these three surfaces are in general \textit{not} manifolds with smooth boundaries, but instead they are manifolds with corners.  Compared to the Schwinger-Keldysh case studied in Section~\ref{sscomb}, the novelty here is that the combinatorial ingredients of $\Delta$ assigned to the six distinct region do not yet generally cover all of $\Sigma$.  We must add yet another region, $\Sigma^{+-\rmm}$, to which we assign all the plaquettes which have adjacent indices of all three types $+$, $-$ and $\rmm$.  The number of such plaquettes will be denoted by $L_{+-\rmm}$.  With the addition of $\Sigma^{+-\rmm}$, each combinatorial element of the cellular decomposition of $\Sigma$ has been accounted for and assigned to exactly one region, and we have defined a partition of $\Sigma$ into seven parts.  

Let us take a closer look at the $\Sigma^{+-\rmm}$ component.  Its combinatorial Euler number will be simply the number of the plaquettes assigned to $\Sigma^{+-\rmm}$,
\be
\chi_{+-\rmm}(\Delta)=L_{+-\rmm}.
\ee
In contrast to the other six regions, which can be topologically complicated with arbitrarily high genus, the topology of $\Sigma^{+-\rmm}$ is quite simple:  Since it contains only plaquettes, and no propagators or vertices of the original ribbon diagram $\Delta$, it consists topologically of a collection of disconnected disks, one for each plaquette.  The entire topology of $\Sigma^{+-\rmm}$ is thus completely fixed in terms of its Euler number $\chi_{+-\rmm}=L_{+-\rmm}$, which simply counts the total number of the disconnected disks.
\begin{figure}[t!]
    \centering
    \includegraphics[width=0.18\textwidth]{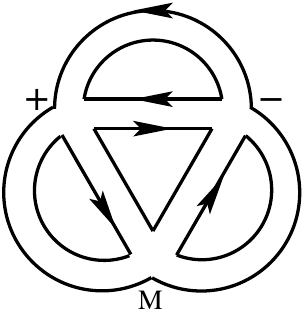}
    \caption{A simple ribbon diagram with a seven-fold decomposition of $\Sigma$.}
    \label{ffseven}
\end{figure}

A simple example of the seven-fold decomposition of $\Sigma$ associated with a ribbon diagram $\Delta$, for which all seven parts of this decomposition are non-empty, is given in Fig.~\ref{ffseven}.  It also provides an example where $\Sigma^{+-}$, $\Sigma^{+\rmm}$ and $\Sigma^{-\rmm}$ are not smooth manifolds, but manifolds with corners, as one can verify by evaluating their Euler numbers.

\begin{figure}[b!]
    \centering
    \includegraphics[width=0.18\textwidth]{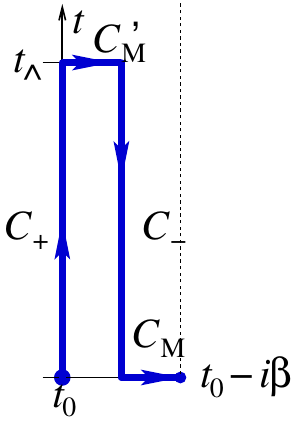}
    \caption{The time contour relevant for thermofield dynamics, as a physically motivated example of a contour with four segments \cite{niesem,das,dast}.}
    \label{ffsk01}
\end{figure}

The decomposition patterns for $\Sigma$ can get even more complicated when one considers time contours with more than three segments.  An extension to such contours is not just a mindless mathematical exercise, as such contours can be physically well-motivated:  For example, the contour relevant for thermofield dynamics has four segments (see Fig.~\ref{ffsk01}).  In the case with $k>3$ segments of the time contour, we introduce an index $i=1,\ldots k$ and iterate our combinatorial construction for $k=3$ from earlier in this section to construct regions $\Sigma^i$, $\Sigma^{ij}$, $\ldots$ $\Sigma^{i_1\ldots i_k}$.  It is best to think of them as antisymmetric in the indices.  The one simplifying feature is that starting from $\Sigma^{ij\ell}$, all higher-order regions consist solely of isolated plaquettes of $\Delta$, and are therefore topologically simple, just as our $\Sigma^{+-\rmm}$ above.  Unfortunately, the $\Sigma^{ij}$'s are again manifolds with corners.

Manifolds with corners are rather awkward, and it would be much preferrable to work only with manifolds with smooth boundaries.  One can avoid using the manifolds with corners in the following way.  First, we define a coarser decomposition of $\Sigma$ into just four parts: Keeping $\sigp$, $\sigm$ and $\Sigma^\rmm$, and assigning all the rest of $\Sigma$ to be the fourth region $\hat\Sigma$. In our example from Fig.~\ref{ffseven}, $\hat\Sigma$ is a sphere with three boundaries. $\hat\Sigma$ can be viewed as a manifold with a smooth boundary, using the same arguments that we used in Section~\ref{sscomb} for $\sigw$.

This four-fold decomposition is rather crude, since it loses track of the regions corresponding to just two segments of the time contour meeting, such as $\Sigma^{+-}$.  We can restore this refinement by the following slight modification of our previous rules:
\begin{itemize}
\item All $+$ vertices, all $G_{++}$ propagators and all the plaquettes which have only $G_{++}$ adjacent propagators define region $\Sigma^+$; analogously for $\Sigma^-$ and $\Sigma^\rmm$.
\item All $G_{+-}$ (and $G_{-+}$) propagators and all the plaquettes that have at least one $G_{+-}$ (or $G_{-+}$) adjacent propagator define region $\tilde\Sigma^{+-}$; analogously for $\tilde\Sigma^{+\rmm}$ and $\tilde\Sigma^{-\rmm}$.
\end{itemize}
Clearly, all the combinatorial ingredients in $\Delta$ have been assigned.  The $\sigp$, $\sigm$ and $\Sigma^\rmm$ regions are defined as before, and they do not share any plaquettes with each other or any other region.  The novelty is in the $\tilde\Sigma$ regions:  They can be interpreted as surfaces with smooth boundaries, but they can now overlap over disks.  Their union is $\hat\Sigma$.  The seventh region $\Sigma^{+-\rmm}$ of the seven-fold decomposition is the collection of all the disks in $\Sigma$ over which at least two of the $\tilde\Sigma$ components overlap.%
\footnote{These new decomposition rules can be extended straightforwardly to the case of time contours with $k$ components.  One defines regions $\Sigma^i$ and $\tilde\Sigma^{ij}$ for $i,j=1,\ldots k$ and $i<j$ in analogy with the $k=3$ case.  They are all manifolds with smooth boundaries. Then $\hat\Sigma=\bigcup\Sigma^{ij}$, and all the higher $\Sigma^{i_1\ldots i_s}$ with $s\geq 3$ correspond to the collection of disks where the appropriate $\tilde\Sigma^{ij}$'s overlap.}
\begin{figure}[t!]
    \centering
    \includegraphics[width=0.17\textwidth]{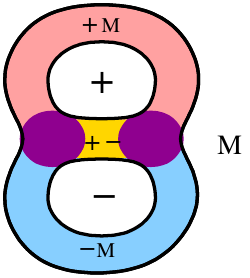}
    \caption{The surface associated with the diagram in Fig.~\ref{ffseven}, and its decomposition into $\sigp$, $\sigm$, $\Sigma^\rmm$, $\tilde\Sigma^{+-}$, $\tilde\Sigma^{+\rmm}$ and $\tilde\Sigma^{-\rmm}$. The three $\tilde\Sigma$'s overlap over two disks.  All components are manifolds with smooth boundaries, and $\hat\Sigma=\tilde\Sigma^{+-}\cup\tilde\Sigma^{+\rmm}\cup\tilde\Sigma^{-\rmm}$ is the sphere with three boundaries.}
    \label{ffsk02}
\end{figure}
In our example from Fig.~\ref{ffseven}, $\tilde\Sigma^{+-}$, $\tilde\Sigma^{+\rmm}$ and $\tilde\Sigma^{-\rmm}$ are all disks, overlapping over two disks, as indicated in Fig.~\ref{ffsk02}.

To conclude this section, we point out that the dual picture of the ribbon diagrams developed in Section~\ref{ssdual} gives an interesting perspective also on the seven-fold decomposition of $\Sigma$ associated with the Kadanoff-Baym contour.  Going from the original ribbon diagram $\Delta$ to its dual diagram $\Delta^\star$ reveals that while $\sigp$, $\sigm$ and $\Sigma^\rmm$ are effectively two-dimensional (since they are built from vertices, lines and plaquettes of $\Delta^\star$), $\Sigma^{+-}$, $\Sigma^{+\rmm}$ and $\Sigma^{-\rmm}$ are effectively one-dimensional, built only from vertices and lines of $\Delta^\star$.  This is reminiscent of what we saw in the triple decomposition on the Schwinger-Keldysh contour in Section~\ref{ssdual}.  In the seven-fold decomposition, this pattern goes one step further, and $\Sigma^{+-\rmm}$ is found to be effectively zero-dimensional, since it is built only from vertices in $\Delta^\star$ and therefore represents just a finite collection of points in this dual picture.

\section{Other generalizations}

Our analysis can be naturally extended from the theory of closed oriented strings to theories containing unoriented and/or open strings.  Since this generalization is straightforward, we will be brief.

\subsection{Unoriented strings}

Until now, we assumed the matrix degrees of freedom to be Hermitian and traceless, in the adjoint representation of the symmetry group $SU(N)$.  We can replace the unitary group $SU(N)$ with another sequence of simple groups that allows a large-$N$ limit -- either orthogonal $SO(N)$ or symplectic $Sp(N)$.  Our story then naturally generalizes and involves unoriented surfaces.

For $SO(N)$ or $Sp(N)$, Feynman rules and their ingredients are essentially the same as in the $U(N)$ case, except that the ribbons now do not carry arrows on their edges,
\be
\vcenter{\hbox{\includegraphics[width=1in]{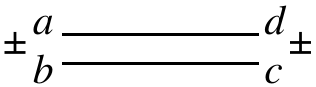}}}\ \ .\nonumber
\ee
The arrows were needed in the $SU(N)$ case to distinguish between the upper indices and the lower indices of $M$, which correspond to inequivalent representations $\mathbf{N}$ and $\overline{\mathbf{N}}$.  In contrast, for $SO(N)$ and $Sp(N)$ the upper and lower indices correspond to the same representation, and can be freely raised and lowered using the invariant quadratic form of $SO(N)$ or $Sp(N)$.  Hence, in Feynman diagrams we no longer have to keep track of the difference between the left and right edge of the ribbons, as reflected by the absence of arrows in the notation.  The matrices $M$ are antisymmetric for $SO(N)$ and symmetric traceless for $Sp(N)$; this difference is immaterial for our arguments, and both cases will lead to the same topological expansion in non-equilibrium string perturbation theory.  (See also Footnote~\ref{ftnt} above for a clarification of the tracelessness condition relevant to the $Sp(N)$ case.)

Similarly, the vertices are 
\bea
\vcenter{\hbox{\includegraphics[width=.75in]{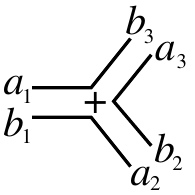}}}\ \ &,&
\ \ \vcenter{\hbox{\includegraphics[width=.75in]{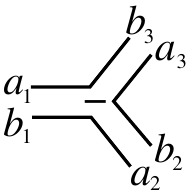}}}\ \ ,\nonumber\\
&&\nonumber\\
\vcenter{\hbox{\includegraphics[width=.8in]{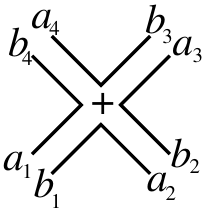}}}\ \ &,&
\ \ \vcenter{\hbox{\includegraphics[width=.8in]{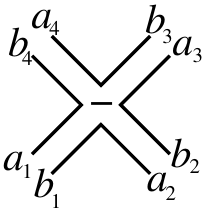}}}\ \ ,\nonumber\\
&\vdots&\ \ .\nonumber
\eea
The dots here stand again for the list of higher $n$-point vertices, which are allowed but kept implicit.

\begin{figure}[t!]
  \centering
    \includegraphics[width=0.2\textwidth]{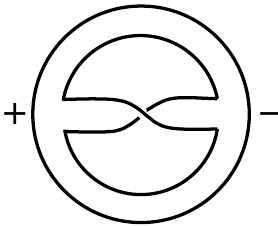}
    \caption{A typical ribbon diagram involving a twist in one of the propagators.  The resulting surface is nonorientable, in this case the projective sphere $\mathbb{R}\mathrm{P}^2$.  In its triple decomposition, $\sigp$ and $\sigm$ are both disks, and $\sigw$ is the sphere with two boundaries and a crosscap.}
    \label{ffsku5}
\end{figure}

Since the edges of the ribbons are no longer oriented, the propagators and vertices can now be connected with an additional twist (see Fig.~\ref{ffsku5}).  The resulting surfaces are then unoriented.

Recall how the classification of closed oriented surfaces extends to the case of closed unoriented surfaces.  Besides the number $h$ of handles, such surfaces $\Sigma$ can also have $c$ crosscaps.%
\footnote{The crosscap is defined by removing a disk from $\Sigma$, which creates an $S^1$ boundary, and then pairwise identifying the opposite points on this boundary; see, \eg , \cite{polchinski}.}
With any nonzero $c$, $\Sigma$ is nonorientable.  In the classification of topologically inequivalent $\Sigma$'s, the two non-negative integers $h$ and $c$ are not independent.  Instead, there is one identity that fully describes their redundancy: $\Sigma$ with $h$ handles and $3+c$ crosscaps (and $b$ boundary components, should those be present) is topologically equivalent to $\Sigma$ with $h+1$ handles and $c+1$ crosscaps (and $b$ boundary components),
\be
\Sigma_{h,c+3,b}=\Sigma_{h+1,c+1,b},
\ee
for all $h=0,1,\ldots$ and $c=0,1,\ldots$ (and $b=0,1,\ldots$).  In equilibrium string theory, the genus expansion is over all inequivalent topologies, classified now by $h$ and $c$ subject to this one identity.  Each surface contributes at order $g_s^{-\chi(\Sigma)}$ in the string coupling, with the Euler number now given by
\be
\chi(\Sigma)=2-2h-c.
\ee
Note that in contrast to the case of closed oriented strings, (i) there are generally several distinct topologies contributing at a given order in $g_s$, and (ii) there are now surfaces that contribute at odd orders in $g_s$.  All this is of course extremely well-undestood in the case of critical string theory \cite{polchinski}.

The results of our analysis for $SU(N)$ in Section~\ref{ssnneq} extend directly to unoriented string theory.  Each surface $\Sigma$ contributing to the non-equilibrium perturbative expansion again exhibits a triple decomposition into the forward region $\sigp$, backward region $\sigm$ and the wedge region $\sigw$, glued together along common boundaries $\p_+\Sigma$ and $\p_-\Sigma$ to form $\Sigma$, the only novelty being that each of the three regions of the triple decomposition can now be orientable or nonorientable.  With this one exception, the story parallels that of Section~\ref{ssnneq}.

\subsection{Coupling to vector degrees of freedom: Open string theory}

Another natural generalization involves the presence of both matrix and vector degrees of freedom, in the adjoint and fundamental representation of one of the large-$N$ sequences $SU(N)$, $SO(N)$ or $Sp(N)$.  This generalization leads to surfaces with boundaries, or in other words, a theory of both closed and open strings.  For simplicity we concentrate on the $SU(N)$ case, which makes the strings oriented; the $SO(N)$ and $Sp(N)$ cases will lead to a description in terms of unoriented closed and open strings.  

Adding the degrees of freedom $\Psi^a$ in the fundamental representation $\mathbf{N}$ (with its conjugate $\bar\Psi_b$ in the anti-fundamental $\overline{\mathbf{N}}$) adds new terms to the action,
\be
S(M,\Psi)=\int \left(\bar\Psi_a\dot\Psi^a+g'\bar\Psi_a M^a{}_b\Psi^b+g''\bar\Psi_aM^a{}_bM^b{}_c\Psi^c+\ldots\right).
\label{eempsi}
\ee
To the Feynman rules for $M$, this will add a propagator for $\Psi$ and new vertices. In equilibrium, the new propagator is the two-point function $\langle\Psi^a\bar\Psi_b\rangle$, which is now denoted by an oriented single line.  When we take the system away from equilibrium, the Schwinger-Keldysh time contour is again that of Fig.~\ref{ffctc}, and it leads to the doubling of fields $\Psi_\pm$, $\bar\Psi_\pm$.  The non-equilibrium propagators thus have each end again labeled by a choice of a $\pm$ sign,%
\footnote{In the quadratic part of (\ref{eempsi}), we again only displayed the term with the time derivative, keeping all the other terms bilinear in $\Psi$ and $\bar\Psi$ (such as masses, terms with spatial derivatives, or with more time derivatives) implicit, to keep the notation simple and to reflect the universality of our arguments.  The propagator (\ref{eevectprop}) of course contains the full information about all such terms.}
\be
\vcenter{\hbox{\includegraphics[width=1in]{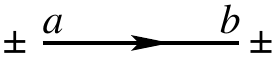}}}\ \ .\nonumber
\label{eevectprop}
\ee
In $S(M.\Psi)$ in (\ref{eempsi}), the ``$\ldots$'' denote interactions with higher powers of $M$.%
\footnote{In the theory of $M$ alone, our interactions were all single-trace, and here we also assume that all the interactions between $\Psi$ and $M$ are of the ``single-trace'' type -- only those monomials that do not factorize into the product of two singlets are admitted.  This in particular implies that the vector degrees of freedom appear quadratically, and all the new vertices have just two single-line ends.  This simplification indeed occurs in various important examples, and in particular mimics the behavior of quarks in QCD.}
Besides the original vertices of the theory with the matrix degrees of freedom $M_\pm$, there are now new vertices,
\bea
\vcenter{\hbox{\includegraphics[width=1in]{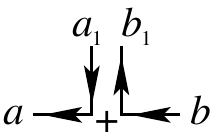}}}\ \ &,&
\ \ \vcenter{\hbox{\includegraphics[width=1in]{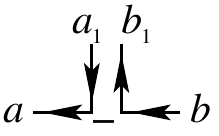}}}\ \ ,\nonumber\\
&&\nonumber\\
\vcenter{\hbox{\includegraphics[width=1in]{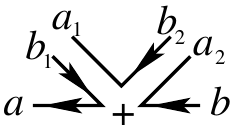}}}\ \ &,&
\ \ \vcenter{\hbox{\includegraphics[width=1in]{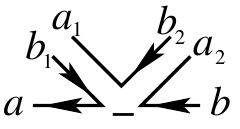}}}\ \ ,\nonumber\\
&\vdots&\nonumber
\eea
which describe the interaction between $M_\pm$, $\Psi_\pm$ and $\bar\Psi_\pm$ on the Schwinger-Keldysh contour.  Our conclusions about the topology of the large-$N$ expansion will be unaffected by whether we choose to think of $\Psi$ as fermions or bosons: Only some signs in individual diagrams change, but the features of the topological expansion remain the same.

Consider for simplicity vacuum Feynman diagrams; the extension to $n$-point correlators is straightforward.  With the vector degrees of freedom $\Psi$ present, any Feynman diagram $\Delta$ will now be associated with a surface $\Sigma$ with boundaries.  The prescription for constructing $\Sigma$ from $\Delta$ is exactly the same as in Section~\ref{ssnstr}.  Following this prescription leaves us with boundaries, each boundary component traced by a closed loop made of the $\Psi\bar\Psi$ propagators.  Thus, the dual string theory contains closed and open oriented strings.

In equilibrium string theory of oriented closed and open strings, the sum over topologies extends over the topologically inequivalent oriented surfaces $\Sigma_{h,b}$ with boundaries, fully classified by the number of handles $h$ and boundaries $b$ which are non-negative integers and without redundancies.  Taking the coupled system of $M$, $\Psi$ away from equilibrium shows that our conclusions from Section~\ref{ssnneq} hold again:  The sum over topologies $\Sigma_{h,b}$ is refined to a sum over triple decompositions $\sigp$, $\sigm$ and $\sigw$ of each $\Sigma_{h,b}$.

At first, it might appear a little awkward that we are supposed to split a surface $\Sigma_{h,b}$ which itself has boundaries, into three regions: Some of the cuts may cut across the boundaries of $\Sigma_{h,b}$.  However, this seemingly intricate issue is easy to deal with, by invoking one of the classic techniques with a proven record in critical string theory in equilibrium \cite{wso,wsiia,sagnotti}, in the context of D-branes and orientifolds:  Treat each worldsheet surface $\Sigma$ with boundaries (and/or crosscaps) as a $\mathbb{Z}_2$ orbifold of a closed oriented surface $\overline\Sigma$, \ie , $\Sigma=\overline{\Sigma}/\mathbb{Z}_2$, with $\mathbb{Z}_2$ an orientation-reversing involution of $\overline\Sigma$.  The boundaries of $\Sigma$ correspond to the lines of $\mathbb{Z}_2$ fixed points of the involution.  The triple decomposition of $\Sigma$ is then simply defined as a triple decomposition of the closed oriented cover $\overline\Sigma$ (in the sense of Section~\ref{ssnneq}), consistent with the $\mathbb{Z}_2$ symmetry.  With this trick, our conclusions of Section~\ref{ssnneq} extend straightforwardly from oriented closed theories to theories with closed and open strings, orientable or nonorientable.

\section{Conclusions}

In this paper, we studied the large-$N$ expansion in non-equilibrium quantum systems with matrix degrees of freedom on the Schwinger-Keldysh time contour, to derive universal features of the perturbative expansion in the dual string theory.  In equilibrium, the standard loop expansion in the powers of the string coupling $g_s$ takes the form of a sum over inequivalent worldsheet topologies $\Sigma$, fully classified (in the case of closed oriented strings) by the number of handles on $\Sigma$.  In non-equilibrium string theory, we found that this topological expansion is further refined: Each surface $\Sigma$ undergoes a triple decomposition into region $\sigp$ on the forward branch of the Schwinger-Keldysh time contour, $\sigm$ on the backward branch of the time contour, and the wedge region $\sigw$ which corresponds to the instant in time where the two branches meet.  Surprisingly, $\sigw$ is itself a topologically two-dimensional region, with arbitrarily complicated topology and its own genus expansion.  The perturbative sum over worldsheet topologies $\Sigma$ now includes a sum over all triple decompositions.

These findings are quite universal, since they follow just from the robust features of the large-$N$ Feynman diagrams, without any assumptions about the (unknown) worldsheet dynamics of the dual theory.  In this sense, we expect that any candidate string-theory dual should consistently reproduce this refined structure of string perturbation theory.  

The next challenge is to find concrete realizations of the refined string perturbation theory in examples where the worldsheet dynamics is known or can be worked out.  At least three natural testing grounds suggest themselves:  One is noncritical string theory in low spacetime dimensions, which is nonperturbatively described by the appropriate continuum limit of matrix models.  Another example, where a lot is known about both sides of the large-$N$/string-theory duality and our ideas can presumably be tested, is the most-studied example of AdS/CFT correspondence, given by $\CN=4$ supersymmetric Yang-Mills theory and its Type IIB superstring $AdS_5\times S^5$ dual.  Finally, critical superstring theory in asymptotically flat spacetimes should also provide interesting tests.  In fact, the insights of this paper may also be relevant for equilibrium superstring perturbation theory, in the context of extending the beautiful methods of Cutkosky rules and Refs.~\cite{veltmanun,diagrr,veltman} for proving unitarity of amplitudes to string theory. These methods have been surprisingly out of reach in the first-quantized approach to string theory (see \cite{ewie} for the relevant discussion), and progress on these issues so far seems to require string field theory \cite{senc}.

We mainly hope that the results of this paper will help to pave the way towards the development of non-equilibrium string theory, enlarging the scope of physical systems that can be described by a string-theory dual.

\acknowledgments
This work has been supported by NSF grants PHY-1820912 and PHY-1521446.

\bibliographystyle{JHEP}
\bibliography{neq}

\providecommand{\href}[2]{#2}\begingroup\raggedright\begin{thebibliography}{10}

\bibitem{glass}
G.~Reggio, P.~Glass and R.~Fricke, \emph{{Koyaanisqatsi, {\rm IRE}}},  1982.

\bibitem{zaanen}
J.~Zaanen, Y.-W. Sun, Y.~Liu and K.~Schalm, \emph{{Holographic Duality in
  Condensed Matter Physics}}. Cambridge Univ. Press, 2015.

\bibitem{hartnoll}
S.~A. Hartnoll, A.~Lucas and S.~Sachdev, \emph{Holographic quantum matter},
  \href{https://arxiv.org/abs/arXiv:1612.07324}{{\ttfamily arXiv:1612.07324}}.

\bibitem{kth}
P.~Ho\v{r}ava, \emph{{Stability of Fermi surfaces and K-theory}},
  \href{https://doi.org/10.1103/PhysRevLett.95.016405}{\emph{Phys. Rev. Lett.}
  {\bfseries 95} (2005) 016405}
  [\href{https://arxiv.org/abs/arXiv:hep-th/0503006}{{\ttfamily
  arXiv:hep-th/0503006}}].

\bibitem{baumann}
D.~Baumann and L.~McAllister, \emph{{Inflation and String Theory}}. Cambridge
  University Press, 2015,
  [\href{https://arxiv.org/abs/arXiv:1404.2601}{{\ttfamily arXiv:1404.2601}}].

\bibitem{birth}
A.~Cappelli, E.~Castellani, F.~Colomo and P.~Di~Vecchia, eds., \emph{{The Birth
  of String Theory}}. Cambridge Univ. Press, Cambridge, UK, 2012.

\bibitem{sonh}
C.~P. Herzog and D.~T. Son, \emph{{Schwinger-Keldysh propagators from AdS/CFT
  correspondence}},
  \href{https://doi.org/10.1088/1126-6708/2003/03/046}{\emph{JHEP} {\bfseries
  03} (2003) 046} [\href{https://arxiv.org/abs/arXiv:hep-th/0212072}{{\ttfamily
  arXiv:hep-th/0212072}}].

\bibitem{kostas1}
K.~Skenderis and B.~C. van Rees, \emph{{Real-time gauge/gravity duality}},
  \href{https://doi.org/10.1103/PhysRevLett.101.081601}{\emph{Phys. Rev. Lett.}
  {\bfseries 101} (2008) 081601}
  [\href{https://arxiv.org/abs/arXiv:0805.0150}{{\ttfamily arXiv:0805.0150}}].

\bibitem{kostas2}
K.~Skenderis and B.~C. van Rees, \emph{{Real-time gauge/gravity duality:
  Prescription, Renormalization and Examples}},
  \href{https://doi.org/10.1088/1126-6708/2009/05/085}{\emph{JHEP} {\bfseries
  05} (2009) 085} [\href{https://arxiv.org/abs/arXiv:0812.2909}{{\ttfamily
  arXiv:0812.2909}}].

\bibitem{haehl1}
F.~M. Haehl, R.~Loganayagam and M.~Rangamani, \emph{{Schwinger-Keldysh
  formalism. Part I: BRST symmetries and superspace}},
  \href{https://doi.org/10.1007/JHEP06(2017)069}{\emph{JHEP} {\bfseries 06}
  (2017) 069} [\href{https://arxiv.org/abs/arXiv:1610.01940}{{\ttfamily
  arXiv:1610.01940}}].

\bibitem{haehl2}
F.~M. Haehl, R.~Loganayagam and M.~Rangamani, \emph{{Schwinger-Keldysh
  formalism. Part II: thermal equivariant cohomology}},
  \href{https://doi.org/10.1007/JHEP06(2017)070}{\emph{JHEP} {\bfseries 06}
  (2017) 070} [\href{https://arxiv.org/abs/arXiv:1610.01941}{{\ttfamily
  arXiv:1610.01941}}].

\bibitem{jandb}
J.~de~Boer, M.~P. Heller and N.~Pinzani-Fokeeva, \emph{{Holographic
  Schwinger-Keldysh effective field theories}},
  \href{https://doi.org/10.1007/JHEP05(2019)188}{\emph{JHEP} {\bfseries 05}
  (2019) 188} [\href{https://arxiv.org/abs/arXiv:1812.06093}{{\ttfamily
  arXiv:1812.06093}}].

\bibitem{ssk}
P.~Ho\v{r}ava and C.~J. Mogni, \emph{{String perturbation theory on the
  Schwinger-Keldysh time contour}},
  \href{https://arxiv.org/abs/arXiv:2009.03940}{{\ttfamily arXiv:2009.03940}}.

\bibitem{th}
G.~'t~Hooft, \emph{{A planar diagram theory for strong interactions}},
  \href{https://doi.org/10.1016/0550-3213(74)90154-0}{\emph{Nucl. Phys. B}
  {\bfseries 72} (1974) 461}.

\bibitem{th2}
G.~'t~Hooft, \emph{{A two-dimensional model for mesons}},
  \href{https://doi.org/10.1016/0550-3213(74)90088-1}{\emph{Nucl. Phys. B}
  {\bfseries 75} (1974) 461}.

\bibitem{thspell}
G.~'t~Hooft, \emph{{Under the Spell of the Gauge Principle, {\rm Chapter 6}}}.
  World Scientific, 1994.

\bibitem{colemann}
S.~R. Coleman, \emph{{1/N}},  in \emph{{17th International School of Subnuclear
  Physics: Pointlike Structures Inside and Outside Hadrons}}, 1980.

\bibitem{coleman}
S.~Coleman, \emph{{Aspects of Symmetry}: {Selected Erice Lectures}}. Cambridge
  University Press, Cambridge, 1985.

\bibitem{juantasi}
J.~M. Maldacena, \emph{{TASI 2003 lectures on AdS/CFT}},  in \emph{{Theoretical
  Advanced Study Institute in Elementary Particle Physics (TASI 2003): Recent
  Trends in String Theory}}, p.~155, 2003,
  \href{https://arxiv.org/abs/arXiv:hep-th/0309246}{{\ttfamily
  arXiv:hep-th/0309246}}.

\bibitem{yaffe}
L.~G. Yaffe, \emph{{Large N limits as classical mechanics}},
  \href{https://doi.org/10.1103/RevModPhys.54.407}{\emph{Rev. Mod. Phys.}
  {\bfseries 54} (1982) 407}.

\bibitem{penner}
R.~C. Penner, \emph{Decorated Teichm\" uller theory}. European Mathematical
  Society, 2012.

\bibitem{juan}
J.~M. Maldacena, \emph{{The Large N limit of superconformal field theories and
  supergravity}}, \href{https://doi.org/10.1023/A:1026654312961}{\emph{Adv.
  Theor. Math. Phys.} {\bfseries 2} (1998) 231}
  [\href{https://arxiv.org/abs/arXiv:hep-th/9711200}{{\ttfamily
  arXiv:hep-th/9711200}}].

\bibitem{schwinger}
J.~S. Schwinger, \emph{{Brownian motion of a quantum oscillator}},
  \href{https://doi.org/10.1063/1.1703727}{\emph{J. Math. Phys.} {\bfseries 2}
  (1961) 407}.

\bibitem{keldysh}
L.~Keldysh, \emph{{Diagram technique for nonequilibrium processes}}, {\emph{Zh.
  Eksp. Teor. Fiz.} {\bfseries 47} (1964) 1515}.

\bibitem{ll}
E.~M. Lifshitz and L.~P. Pitaevskii, \emph{{Landau and Lifshitz Course of
  Theoretical Physics, Volume 10: Physical Kinetics}}. Pergamon Press, 1981.

\bibitem{rammers}
J.~Rammer and H.~Smith, \emph{{Quantum field-theoretical methods in transport
  theory of metals}},
  \href{https://doi.org/10.1103/RevModPhys.58.323}{\emph{Rev. Mod. Phys.}
  {\bfseries 58} (1986) 323}.

\bibitem{rammer}
J.~Rammer, \emph{{Quantum field theory of non-equilibrium states}}. Cambridge
  University Press, Cambridge, 2007.

\bibitem{mahan}
G.~D. Mahan, \emph{{Many-Particle Physics, Third Edition}}. Plenum Press, 2000.

\bibitem{chou}
K.-c. Chou, Z.-b. Su, B.-l. Hao and L.~Yu, \emph{{Equilibrium and
  nonequilibrium formalisms made unified}},
  \href{https://doi.org/10.1016/0370-1573(85)90136-X}{\emph{Phys. Rept.}
  {\bfseries 118} (1985) 1}.

\bibitem{lebellac}
M.~Le~Bellac, \emph{{Thermal Field Theory}}. Cambridge University Press,
  Cambridge, 1996.

\bibitem{das}
A.~K. Das, \emph{{Finite Temperature Field Theory}}. World Scientific, New
  York, 1997.

\bibitem{dast}
A.~K. Das, \emph{{Topics in finite temperature field theory}},
  \href{https://arxiv.org/abs/arXiv:hep-ph/0004125}{{\ttfamily
  arXiv:hep-ph/0004125}}.

\bibitem{kamenev}
A.~Kamenev, \emph{{Field Theory of Non-Equilibrium Systems}}. Cambridge
  University Press, Cambridge, 2011.

\bibitem{svl}
G.~Stefanucci and R.~van Leeuwen, \emph{{Nonequilibrium Many-Body Theory of
  Quantum Systems}}. Cambridge University Press, Cambridge, 2013.

\bibitem{gelis}
F.~Gelis, \emph{{Quantum Field Theory: From Basics to Modern Topics}}.
  Cambridge University Press, 2019.

\bibitem{vilkovisky}
G.~Vilkovisky, \emph{{Expectation values and vacuum currents of quantum
  fields}}, {\emph{Lect. Notes Phys.} {\bfseries 737} (2008) 729}
  [\href{https://arxiv.org/abs/arXiv:0712.3379}{{\ttfamily arXiv:0712.3379}}].

\bibitem{spicka1}
V.~\v{S}pi\v{c}ka, B.~Velick\'{y} and A.~Kalvov\'{a}, \emph{{Long and short
  time quantum dynamics: I. Between Green's functions and transport
  equations}}, {\emph{Physica} {\bfseries E 29} (2005) 154}.

\bibitem{spicka2}
V.~\v{S}pi\v{c}ka, B.~Velick\'{y} and A.~Kalvov\'{a}, \emph{{Long and short
  time quantum dynamics: II. Kinetic regime}}, {\emph{Physica} {\bfseries E 29}
  (2005) 175}.

\bibitem{spicka3}
V.~\v{S}pi\v{c}ka, B.~Velick\'{y} and A.~Kalvov\'{a}, \emph{{Long and short
  time quantum dynamics: III. Transients}}, {\emph{Physica} {\bfseries E 29}
  (2005) 196}.

\bibitem{spicka}
V.~\v{S}pi\v{c}ka, B.~Velick\'{y} and A.~Kalvov\'{a}, \emph{{Electron systems
  out of equilibrium: Nonequilibrium Green's function approach}},
  \href{https://doi.org/10.1142/S0217979214300138}{\emph{Int. J. Mod. Phys.}
  {\bfseries B 28} (2014) 1430013}.

\bibitem{weinberg1}
S.~Weinberg, \emph{{Quantum contributions to cosmological correlations}},
  \href{https://doi.org/10.1103/PhysRevD.72.043514}{\emph{Phys. Rev. D}
  {\bfseries 72} (2005) 043514}
  [\href{https://arxiv.org/abs/arXiv:hep-th/0506236}{{\ttfamily
  arXiv:hep-th/0506236}}].

\bibitem{weinberg2}
S.~Weinberg, \emph{{Quantum contributions to cosmological correlations. II. Can
  these corrections become large?}},
  \href{https://doi.org/10.1103/PhysRevD.74.023508}{\emph{Phys. Rev. D}
  {\bfseries 74} (2006) 023508}
  [\href{https://arxiv.org/abs/arXiv:hep-th/0605244}{{\ttfamily
  arXiv:hep-th/0605244}}].

\bibitem{weinberg3}
S.~Weinberg, \emph{{Effective field theory for inflation}},
  \href{https://doi.org/10.1103/PhysRevD.77.123541}{\emph{Phys. Rev. D}
  {\bfseries 77} (2008) 123541}
  [\href{https://arxiv.org/abs/arXiv:0804.4291}{{\ttfamily arXiv:0804.4291}}].

\bibitem{diagrr}
G.~'t~Hooft and M.~Veltman, \emph{{Diagrammar}},
  \href{https://doi.org/10.1007/978-1-4684-2826-1\_5}{\emph{NATO Sci. Ser. B}
  {\bfseries 4} (1974) 177}.

\bibitem{veltmanun}
M.~Veltman, \emph{{Unitarity and causality in a renormalizable field theory
  with unstable particles}},
  \href{https://doi.org/10.1016/S0031-8914(63)80277-3}{\emph{Physica}
  {\bfseries 29} (1963) 186}.

\bibitem{veltman}
M.~Veltman, \emph{{Diagrammatica: The Path to Feynman rules}}. Cambridge
  University Press, 2012.

\bibitem{leila}
L.~Schneps, \emph{{The Grothendieck Theory of Dessins d'Enfants, \textrm{LMS
  Lecture Note Series} \textbf{200}}}. Cambridge University Press, 1994.

\bibitem{graphs}
S.~K. Lando and A.~K. Zvonkin, \emph{{Graphs on Surfaces and Their
  Applications}}. Springer-Verlag, 2004.

\bibitem{guillot}
P.~Guillot, \emph{{An elementary approach to dessins d'enfants and the
  Grothendieck-Teichmüller group}},
  \href{https://arxiv.org/abs/arXiv:1309.1968[math.GR]}{{\ttfamily
  arXiv:1309.1968[math.GR]}}.

\bibitem{dessins}
G.~A. Jones and J.~Wolfart, \emph{{Dessins d'Enfants on Riemann Surfaces}}.
  Springer, 2016.

\bibitem{galois}
N.~A'Campo, L.~Ji and A.~Papadopoulos, \emph{{Actions of the absolute Galois
  group}},  \href{https://arxiv.org/abs/arXiv:1603.03387}{{\ttfamily
  arXiv:1603.03387}}.

\bibitem{dessbe}
S.~Bose, J.~Gundry and Y.-H. He, \emph{{Gauge theories and dessins d`enfants:
  beyond the torus}},
  \href{https://doi.org/10.1007/JHEP01(2015)135}{\emph{JHEP} {\bfseries 01}
  (2015) 135} [\href{https://arxiv.org/abs/arXiv:1410.2227}{{\ttfamily
  arXiv:1410.2227}}].

\bibitem{desscy}
Y.-H. He, J.~McKay and J.~Read, \emph{{Modular subgroups, dessins d'enfants and
  elliptic K3 surfaces}},
  \href{https://doi.org/10.1112/S1461157013000119}{\emph{J. Comput. Math.}
  {\bfseries 16} (2013) 271}
  [\href{https://arxiv.org/abs/arXiv:1211.1931}{{\ttfamily arXiv:1211.1931}}].

\bibitem{desscyy}
Y.-H. He, \emph{{Calabi-Yau varieties: from quiver representations to dessins
  d'enfants}},  \href{https://arxiv.org/abs/arXiv:1611.09398}{{\ttfamily
  arXiv:1611.09398}}.

\bibitem{stable}
E.~Arbarello, M.~Cornalba and P.~A. Griffiths, \emph{{Geometry of Algebraic
  Curves, Volume II}}. Springer, 2011.

\bibitem{kadb}
L.~P. Kadanoff and G.~Baym, \emph{{Quantum Statistical Mechanics: Green's
  Function Methods in Equilibrium and Nonequilibrium Problems}}. W.A. Benjamin,
  New York, 1962.

\bibitem{niesem}
A.~Niemi and G.~Semenoff, \emph{{Finite temperature quantum field theory in
  Minkowski space}},
  \href{https://doi.org/10.1016/0003-4916(84)90082-4}{\emph{Annals Phys.}
  {\bfseries 152} (1984) 105}.

\bibitem{polchinski}
J.~Polchinski, \emph{{String Theory. Vol. 1: An introduction to the bosonic
  string}}. Cambridge University Press, 1998.

\bibitem{wso}
P.~Ho\v{r}ava, \emph{{Strings on worldsheet orbifolds}},
  \href{https://doi.org/10.1016/0550-3213(89)90279-4}{\emph{Nucl. Phys. B}
  {\bfseries 327} (1989) 461}.

\bibitem{wsiia}
P.~Ho\v{r}ava, \emph{{Background duality of open string models}},
  \href{https://doi.org/10.1016/0370-2693(89)90209-8}{\emph{Phys. Lett. B}
  {\bfseries 231} (1989) 251}.

\bibitem{sagnotti}
A.~Sagnotti, \emph{{Open strings and their symmetry groups}},  in \emph{{NATO
  Advanced Summer Institute on Nonperturbative Quantum Field Theory (Cargese
  Summer Institute)}}, p.~0521, 1987,
  \href{https://arxiv.org/abs/arXiv:hep-th/0208020}{{\ttfamily
  arXiv:hep-th/0208020}}.

\bibitem{ewie}
E.~Witten, \emph{{The Feynman $i \epsilon$ in string theory}},
  \href{https://doi.org/10.1007/JHEP04(2015)055}{\emph{JHEP} {\bfseries 04}
  (2015) 055} [\href{https://arxiv.org/abs/arXiv:1307.5124}{{\ttfamily
  arXiv:1307.5124}}].

\bibitem{senc}
R.~Pius and A.~Sen, \emph{{Cutkosky rules for superstring field theory}},
  \href{https://doi.org/10.1007/JHEP10(2016)024}{\emph{JHEP} {\bfseries 10}
  (2016) 024} [\href{https://arxiv.org/abs/arXiv:1604.01783}{{\ttfamily
  arXiv:1604.01783}}].

\end{thebibliography}\endgroup
\end{document}